\documentclass{aa}
\usepackage{graphicx}
\usepackage{txfonts}
\usepackage{lscape}
\usepackage{longtable}
\usepackage[switch]{lineno}
\usepackage{xcolor}
\usepackage{comment}

\usepackage[allcolors=blue]{hyperref}
\begin{document}

   \title{Extragalactic stellar tidal streams: observations meet simulation}

   \author{Juan Mir\'{o}-Carretero \inst{1,2}, Mar\'{i}a A. G\'{o}mez-Flechoso \inst{1,3}, David Mart\' \i nez-Delgado\inst{4,5,6}\thanks{ARAID Fellow}, Andrew P. Cooper \inst{7,8}, Santi Roca-F\`abrega \inst{9}, Mohammad Akhlaghi \inst{4}, Annalisa Pillepich \inst{10}, Konrad Kuijken \inst{2}, Denis Erkal \inst{11}, Tobias Buck \inst{12,13}, Wojciech A. Hellwing \inst{14}, Sownak Bose \inst{15}, Giuseppe Donatiello \inst{16}, Carlos S. Frenk \inst{15}
   }

    \institute{Departamento de F{\'\i}sica de la Tierra y Astrof{\'\i}sica, Universidad Complutense de Madrid, Plaza de las Ciencias 2, E-28040 Madrid, Spain 
    \and
    Leiden Observatory, Leiden University, P.O. Box 9513, 2300 RA Leiden, The Netherlands
    \and
    Instituto de F\'isica de Part\'iculas y del Cosmos (IPARCOS), Fac. CC. F\'isicas, Universidad Complutense de Madrid, Plaza de las Ciencias, 1, E-28040 Madrid, Spain
    \and
    Centro de Estudios de F\'isica del Cosmos de Arag\'on (CEFCA), Unidad Asociada al CSIC, Plaza San Juan 1, 44001 Teruel, Spain
    \and
     ARAID Foundation, Avda. de Ranillas, 1-D, E-50018 Zaragoza, Spain
     \and
    Instituto de Astrof\'isica de Andaluc\'ia, CSIC, Glorieta de la Astronom\'\i a, E-18080, Granada, Spain    
    \and
    Institute of Astronomy and Department of Physics, National Tsing Hua University, Kuang Fu Rd. Sec. 2, Hsinchu 30013, Taiwan
    \and
    Center for Informatics and Computation in Astronomy, National Tsing Hua University, Kuang Fu Rd. Sec. 2, Hsinchu 30013, Taiwan
    \and 
    Lund Observatory, Division of Astrophysics, Department of Physics, Lund University, Box 43, SE-221 00 Lund, Sweden
    \and
    Max Planck Institut f{\"u}r Astronomie, K{\"o}nigstuhl 17, D-69117 Heidelberg, Germany
    \and 
    Department of Physics, University of Surrey, Guildford GU2 7XH, UK
   \and
    Universität Heidelberg, Interdisziplinäres Zentrum für Wissenschaftliches Rechnen, Im Neuenheimer Feld 205, D-69120 Heidelberg, Germany
    \and
    Universität Heidelberg, Zentrum für Astronomie, Institut für Theoretische Astrophysik, Albert-Ueberle-Straße 2, D-69120 Heidelberg, Germany
    \and
    Center for Theoretical Physics, Polish Academy of Sciences, Al. Lotnik ow 32/46, 02-668 Warsaw, Poland
    \and
    Institute for Computational Cosmology, Department of Physics, Durham University, South Road, Durham DH13LE, United Kingdom
    \and
    UAI - Unione Astrofili Italiani /P.I. Sezione Nazionale di Ricerca Profondo Cielo, 72024 Oria, Italy
}
\titlerunning{Extragalactic Stellar Tidal Streams: Observations meet Simulation}
\authorrunning{Mir\'{o}-Carretero et al.}

% \abstract{}{}{}{}{} 
% 5 {} token are mandatory
 
  \abstract
  % context heading (optional)
  % {} leave it empty if necessary  
   {According to the well established hierarchical framework for galaxy evolution, galaxies grow through mergers with other galaxies and the $\Lambda$CDM cosmological model predicts that the stellar halos of massive galaxies are rich in remnants from minor mergers. The Stellar Streams Legacy Survey (SSLS) has provided a first release of a catalogue with a statistically significant sample of stellar streams in the local Universe, detected in deep images from DESI Legacy Surveys and the Dark Energy Survey (DES).}
  % aims heading (mandatory) 
   {The main objective is to compare observations of stellar tidal streams from the SSLS catalogue with predictions from state-of-the-art cosmological simulations regarding their abundance, up to a redshift z < 0.02, according to the $\Lambda$CDM model.}
  % methods heading (mandatory)
   {In particular, we use the outcome of the cosmological simulations Copernicus Complexio, TNG50 of the IllustrisTNG project, and Auriga to generate mock images of nearby halos and search for stellar streams. We compare the stream frequency and characteristics found in these images, as well as the results of a photometric analysis of the simulations data, with DES observations.}
  % results heading (mandatory)
   {We find generally good agreement between the real images and the simulated ones regarding frequency and photometry of streams, while the stream morphology is somewhat different between observations and simulations, and between simulations themselves. By varying the sky background of the synthetic images to emulate different surface brightness limit levels, we also obtain predictions for the detection rate of stellar tidal streams up to a surface brightness limit of 35 mag arcsec$^{-2}$.}
  % conclusions heading (optional), leave it empty if necessary } 
  {The cosmological simulations predict that with an instrument such as the one used in the DES, it would be necessary to reach a surface brightness limit of 32 mag arcsec$^{-2}$ in the $r$-band to achieve a frequency of up to $\sim$ 70\% in the detection of stellar tidal streams around galaxies in the redshift range considered here.}

   \maketitle
%
%-------------------------------------------------------------------
\section{Introduction}
\label{sec:introduction}

According to the well established hierarchical framework for galaxy evolution, galaxies grow through mergers with other galaxies. These mergers can be major mergers, when the merging galaxies are of similar stellar mass; a mass ratio > $1/3$ is a generally accepted threshold, see e.g. \citet{newberg2016}, and minor mergers, when the host galaxy accretes a dwarf galaxy in its halo. 
The Lambda Cold Dark Matter ($\Lambda$CDM) cosmological model predicts that the stellar halos of massive galaxies ($\log_{10} M_{\star}/M_{\odot} \gtrsim 9$)  are rich in remnants from minor mergers that, in the local Universe, on the basis of observations and simulations, are expected to be more frequent than major mergers \citep{guo2008,jackson2022}. 
 
\begin{figure*}
\centering
\includegraphics[width=1.0\textwidth]{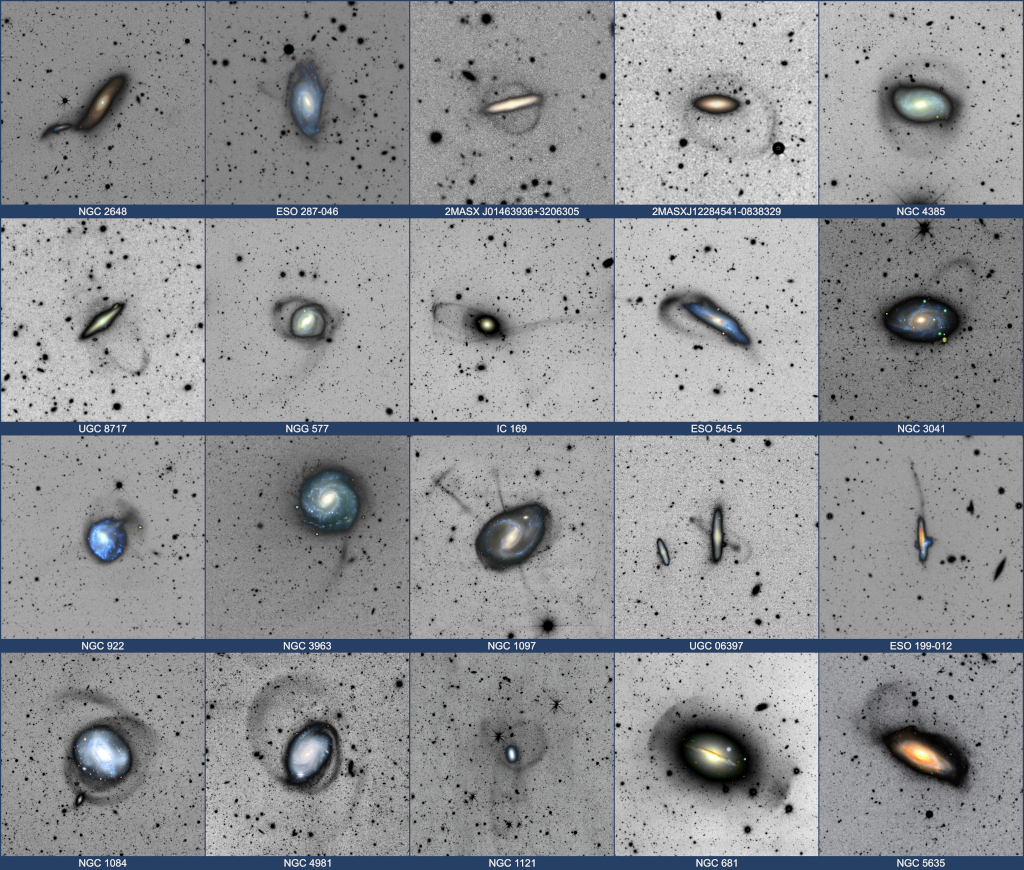}
  \caption{Examples of stellar stream images from the \textit{Stellar Streams Legacy Survey} \citep{martinez-delgado2024}. The cutout angular sizes vary between 3x3 and 10x10 arcmin.} 
  \label{fig-streams}
\end{figure*}

The detection and study of extragalactic stellar tidal streams contributes to augment the stream census, mostly built so far from streams detected in the Milky Way (MW) and Local Volume \citep{belokurov2006,martinez-delgado2010,hood2018,shipp2018,ferguson2022,li2022}, and helps us to contrast their frequency and characteristics with the predictions of the $\Lambda$CDM model on a statistically sound basis (the Local Volume is considered here to be a spherical region with a radius of 11 Mpc around the MW  or up to a radial velocity of redshift of z < 0.002). 
%This implies to look
This motivates the search  for streams beyond the Local Volume, and up to a distance for which surveys are available with the required depth \citep[for the most part streams are expected to be fainter than $\sim$ 30 mag arcsec$^{-2}$, e.g.][]{johnston2008}. %and therefore, 
Due to their low surface brightness  
%such streams present a reduced number of observables with respect to the tidal 
much less data can be gathered for each individual stream in a survey of distant hosts, in comparison to the  streams in our galaxy or in the Local Volume. 
%Among others, no 6D information 
% is available of 
In particular, star-by-star photometric and kinematic measurements are usually not possible for  streams at these distances.
%Also, the stars forming the stream are not resolved. 

A number of relevant surveys of tidal features beyond the Local Volume have been reported in the last decade: \citet{duc2015,morales2018,bilek2020,sola2022,martinez-delgado2023b,miro-carretero2023,giri2023,rutherford2024,skryabina2024,miro-carretero2024}. \citet{skryabina2024} present the results of visual inspection of a sample of 838 edge-on  galaxies using images from three surveys: SDSS Strip-82, Subaru HSC and DESI (DECals, MzLS, BASS). This study, as our present work, was motivated to construct a deep photometric sample and obtain better statistics of tidal structures in the local Universe in order to compare with cosmological simulations. The definition of tidal features used in that study includes also disc deformations and tidal tails, typical of major mergers. Their results will be discussed further in Section \ref{sec:discussion}.

%\apc{See overleaf comments on this paragraph}
Simulations are required to interpret the observations and to infer the 
physics that determines the origin and evolution of streams. 
The commonly used analytical/semi-empirical methods to study stellar streams formation in the local Universe cannot be used when the observational data is scarce and when the central system’s mass distribution evolves with time. So, to understand the formation of the unresolved stellar streams at large distances, accounting for the full temporal evolution, cosmological simulations are needed. However, cosmological simulations also have disadvantages.
They cannot reach the high spatial and mass resolution of bespoke models. Furthermore, the impact of uncertainties in sub-grid models of baryonic astrophysics (e.g.\ star formation and supernova feedback) is poorly known on the very small scales probed by tidal streams.
%as the analytical and semi-empirical models and the dependence of stream formation on the numerics (resolution, code and sub-grid physics) is not well understood.
 
 %The analytical methods used to modelling the streams do not work for extragalactic streams.
 %\textbf{Sanderson}. 

\begin{table*}
\centering
%\small
{\caption{Cosmology parameters used in the simulations. For COCO, $L$ is the radius of high-resolution region, and the value of $\Omega_{b,0}$ is that assumed when generating initial conditions and in the semi-analytic model. Since COCO is a collisionless $N$-body simulation, only a single particle species is used in the high-resolution region, the mass of which (given in the column $m_{DM}$) includes the contribution of both dark matter and baryons to the total matter density. The particle-tagging treatment of stellar mass in COCO is described in the text.}
\label{tab:cosmology}

\begin{tabular}{lccccccccc}

  & $L$     & $m_{DM}$           & $m_{\star}$   & $\Omega_{\Lambda,0}$ &  $\Omega_{m,0}$ & $\Omega_{b,0}$ & $\sigma_{8}$ &  $n_{S}$ &  $h$  \\
            &  Mpc  & $M_{\odot}$ & $M_{\odot}$   &                      &                 &                &            &          &       \\

\hline\hline

COCO	&	17.4	$h^{-1}$    & $1.35 \times 10^{5} h^{-1}$	&	                  & 0.728  & 0.272  & 0.04455 & 0.81   & 0.967  & 0.704 \\
TNG50	&	$50$	           & $4.5 \times 10^{5}$	        & $8.5 \times 10^{4}$ &	0.6911 & 0.3089 & 0.0486  & 0.8159 & 0.9667 & 0.6774 \\
AURIGA	&	$100\, h^{-1}\,$   &	$3 \times 10^{5}$           & $5 \times 10^{4}$   & 0.693  & 0.307  & 0.048   & 0.811  & 0.965  & 0.6777 \\

\hline

\end{tabular}
}
\end{table*}
 
 %Cosmological simulations, while generally costly, can provide a good representation of stream formation and evolution up to a high redshift. 
 Cosmological simulations nevertheless provide a powerful means  not only to understand the origin and evolution of streams as observed in the surveys above, but also 
 to predict their photometric characteristics.
 State-of-the-art models are now detailed enough to be constrained by stream observations, and can also inform assessments of the design and completeness of the observations themselves.
 %and thereby facilitate the assessment of the capabilities needed for their detection.
 We already know that future surveys will need to be able to produce deeper images than available today, if new extra galactic streams, presumed to exist in great numbers, are to be discovered. 
 %But
 However, it remains unclear what %is the 
 critical image depth
 is
 needed to significantly increase %today's detection rate
 the number of known streams.
 %remains unclear. 
 This is important 
 %to conceive and plan future 
 to motivate and plan for
 surveys such as ESA's space mission Euclid \citep{euclid2018,hunt2024} and the Vera C. Rubin Observatory's Legacy Survey of Space and Time \citep[LSST,][]{martin2022,khalid2024}. 
 In the context of minor mergers, predictions for streams and their progenitor galaxies are close to the limit of the capabilities of current large-volume cosmological simulations, and the robustness of those predictions has not yet been explored in detail. The use of cosmological simulations to plan for and interpret new surveys therefore has to proceed in tandem with their validation against existing extragalactic stream data, mostly at surface brightness limits brighter than $\sim$ 29 mag arcsec$^{-2}$ \citep{shipp2023}.
 % The cosmological simulations need to be validated for our purpose, and in absence of observations at fainter surface brightness limits than $\sim$ 29 mag arcsec$^{-2}$), the only method at hand is to draw from the comparison between state-of-the-art cosmological simulations available today. 

The use of cosmological simulations to study tidal features has also increased significantly in the recent times \citep{martinez-delgado2019,mancillas2019,martin2022,vera-casanova2022,valenzuela2024, khalid2024}
Relevant work on stream detectability using mock-images from cosmological simulations is reported in \citet{vera-casanova2022}. The authors have inspected surface brightness maps generated from 30 Auriga project simulations \citep{grand2017} of MW-like galaxies looking for the brightest streams. They report that no streams have been detected in images with a surface brightness limit brighter than 25 mag arcsec$^{-2}$. Their stream detection frequency increases significantly between 28 and 29 mag arcsec$^{-2}$. They 
%go on to assert 
find a correlation between infall time and infall mass of the 
%streams
stream progenitors, such that more massive progenitors tend to be accreted at later times.
% We highlight that in this work we only consider as streams low surface brightness features that are of an accreted origin.
 
\citet{martin2022} report on a theoretical investigation of the extended diffuse light around galaxies and galaxy groups by visually inspecting mock-images produced using the NEWHORIZON cosmological simulations. This is carried out on a sample of 37 simulated objects at redshifts $z = 0.2, 0.4, 0.6$ and $0.8$, spanning a stellar mass range of $10^{9.5} < M_{\star} < 10^{11.5} \mathrm{M}_{\odot}$.
Through production of surface brightness maps at different surface brightness limits, they predict the fraction of tidal features that can be expected to be detected at different limiting surface brightnesses. 
\citet{khalid2024}
%, the results of identification and classification 
identified and classified
%of 
tidal features in LSST-like mock-images from 
%cosmological simulations 
%are reported. 
four sets of hydrodynamical cosmological simulations:
%are used (\textsc{NewHorizons}, 
\textsc{EAGLE}, \textsc{IllustrisTNG} and \textsc{Magneticum}). 
%\textit{Tidal Features} 
These features
comprise streams/tails, shells, plumes or asymmetric stellar halos and double nuclei, and as such do not distinguish between minor and major mergers as the origin of the such features. The results of this previous work and the one presented in the preceding paragraphs will be discussed in more detail in Section \ref{subsec:comparisonpreviouswork}.

As in this work, the works by \citet{martin2022} and \citet{khalid2024} rely on visual inspection of mock-images from cosmological simulations. However 
%the references cited 
they
focus on the detection of \textit{tidal features} and \textit{tidal tails}, while we focus our analysis on the detection and characterisation of remnants of minor mergers, low surface brightness features that are of an accreted origin. 
%They constitute a subset of \textit{tidal features} that we refer to in this paper by the name \textit{tidal streams}, following on the terminology used in \citet{martinez-delgado2023b}, and independently of the observable morphology (e.g. circles, umbrellas/shells or plumes). 
One important conclusion of \citet{martin2022}, with which we concur, is that a higher level of domain knowledge is required to perform robust visual classifications of tidal features (more so than to separate spiral and elliptical galaxies, for example). This work follows 
%on previous work by the authors 
our previous surveys  
to detect stellar streams in %observational 
images from the DESI Legacy Surveys \citep{martinez-delgado2023,miro-carretero2023,miro-carretero2024}, thus, the inspection of mock images in this paper benefits from our experience gathered from working with comparable observational images. 
 
In this work we use the term \textit{stellar tidal streams} to refer to the remnants of minor mergers, in line with the nomenclature used in \citet{martinez-delgado2023}\footnote{\textit{great circles}, hereinafter referred to as circles, are streams that result from satellites along mildly eccentric orbits, with an arc-like shape, sometimes featuring complete loops around the host, but in most cases (in our sample) seen as covering only a small part of a loop; \textit{umbrellas}, structures often appearing on both sides of the host galaxy, displaying an elongated shaft ending in the form of a \textit{shell} (sometimes only the shells are visible) resulting from satellites that were on more eccentric, radial orbits; \textit{giant plumes}, hereinafter referred to as plumes, structures appearing to shoot out of the host, generally for quite a long distance}. 
Our work focuses on stellar tidal streams in the local Universe up to a distance of 100 Mpc (redshift $z < 0.02$). 
We consider as stellar tidal streams only those low surface
brightness (LSB) features that are of an accreted origin, whatever their apparent morphology (shells, circles, plumes etc.); as we will discuss later in the paper, the apparent morphology is strongly dependent on the line of sight of the observation. We can broadly characterise stellar tidal streams as LSB structures in the halo of galaxies, at distances between $\sim$ 20 and 120 kpc from the host centre and with surface brightness fainter than $\sim$ 25 mag arcsec$^{-2}$ for the most part.
Stellar tidal streams are thus a particular case of LSB structures and are significantly (several mag arcsec$^{-2}$) fainter than tidal tails, another type of LSB feature
resulting primarily from major mergers \citep{toomre1972}.

The results from observations of the Dark energy Survey (DES) presented in \citet{miro-carretero2024} allow for a direct, quantitative comparison of the abundance and characteristics of stellar tidal streams in the local Universe with the predictions from state-of-the-art cosmological simulations based on the $\Lambda$CDM paradigm. 
%This includes the comparison of the obtained statistical analysis results 
In particular, we can compare statistics derived from the observed stream population (for example, the number of stream detections at at given surface brightness limit, or the distribution of photometric observables for detected streams)
with those predicted by cosmological simulations, as well as the comparison of the measured photometry.   
To do this,
%we use the output of selected cosmological simulations as the best-estimate prediction for the formation of stellar streams that can be observed at redshift z < 0.02. 
we obtain predictions of stream formation from three cosmological simulations: Copernicus Complexio \citep[COCO,][]{hellwing2016}, TNG50 \citep{pillepich2019,nelson2019a,nelson2019b} and Auriga \citep{grand2017}. 
%In Section \ref{sec:cosmosimulations} we provide an overview of these simulations. 

We have carried out this work in the context of the {\it Stellar Stream Legacy Survey} \citep[SSLS, see ][]{martinez-delgado2023b}, whose main objective is to perform a systematic survey of stellar tidal streams in a parent galaxy sample of $\sim$ 3200 nearby galaxies using images from the recently completed DESI Legacy Survey imaging surveys. Examples of stellar streams detected in the \textit{SSLS} can be seen in Figure \ref{fig-streams}. %The first batch of the survey has yielded a catalogue of streams presented in 
A catalog of streams from the first batch of galaxies in this survey is presented in \citet{miro-carretero2024}. 
 
In this paper, we
%present a comparison of the results of searching and characterising  stellar streams in an image sample of the DES survey 
compare the stellar streams in a sample of galaxies observed by the DES survey
with 
%the results of carrying our this activity in a 
streams in mock images 
%from three sets of cosmological simulations: COCO, TNG50 and Auriga. 
derived from the simulations listed above, for a matched sample of hosts.
We compare the detection frequency and photometric characteristics measured in both samples
%, observations from the DES survey and mock images from cosmological simulations, 
and discuss the results. 
%of such comparison.
In Section \ref{sec:cosmosimulations} we introduce the main characteristics of the 
%selected 
cosmological simulations. The selection of the halos from the simulations to be analysed is presented in Section \ref{sec:haloselection}. Section \ref{sec:mockimages} is devoted to the process of generating mock images. In Section \ref{sec:streamdetection} we present the predicted detectability of streams at different surface brightness limits. In Section \ref{sec:streamphotometry} we  present the results of the photometry measurements in the mock images. 
The results of the comparison are discussed in Section \ref{sec:discussion} and the summary, conclusions and outlook are given in Section \ref{sec:conclusions}.

\section{Cosmological Simulations}
\label{sec:cosmosimulations}

%It is not the goal of this paper to describe in detail the characteristics of cosmological simulations. 
This work makes use of several different cosmological simulations. An overview of the available cosmological simulations, as well as the underlying tools and modelling paradigms, can be found in \citet{vogelsberger2020} and as part of the AGORA collaboration \citep{roca-fabrega2024}. For our work, we have selected three simulation sets, each belonging to one of the broad categories in which the cosmological simulations are classified at the highest level:

\begin{itemize}

\item \textit{Volume simulations} produce large, statistically complete samples of galaxies 
 but typically do not resolve spatial scales smaller than $\sim100$~pc. Physical processes on scales smaller than the explicit hydrodynamical scheme, such as star formation and feedback, are incorporated via semi-analytical `sub-grid' models.

 \item \textit{Zoom-in simulations} produce smaller samples of galaxies with a higher spatial and mass resolution and thereby model baryonic processes on smaller scales.

 \item \textit{Semi-analytical simulations} are the result of a combination of numerical dark matter-only simulations, and analytic models for the prescription of baryonic physics. They are computationally much more efficient than the above categories, at the cost of self-consistency in the dynamics of the baryonic component.
 
\end{itemize}

For our comparison with the observational data we have chosen one state-of-the-art simulation suite of each of the types listed above.

\subsection{Copernicus Complexio}
\label{sec:coco}

The \textit{Copernicus Complexio} \citep[COCO.][]{hellwing2016} is a $\Lambda$CDM cosmological $N$-body simulation post-processed with a semi-analytic galaxy formation model and the `Stellar Tags in N-Body Galaxy Simulations' (STINGS) particle tagging technique \citep{cooper2010,cooper2017}.
%and were carried out by the Durham University. 
COCO provides both high mass resolution %($1.6 \times 10^5$ M$_{\odot}$/particle) 
and an approximate analogue of the Local Volume (a high-resolution spherical region of radius $\sim$25 Mpc, with density slightly lower than the cosmic mean, embedded in a lower-resolution box of 100 Mpc/side). 
%The results of cosmological simulations with COCO exist for the $\Lambda$CDM model paradigm and have been used as input to this task.
The Galform semi-analytic model of \citet{lacey2016} was used to predict the evolution of the baryonic component in each dark matter halo. This model is calibrated to a range of low and high-redshift observables, including optical and near-IR luminosity functions, the HI mass function, and the relationship between the masses of bulges mass and central supermassive black holes. The 6-dimensional phase space of each single-age stellar population formed in the model is mapped to an individual subset of dark matter particles using the STINGS technique. These models will be presented in detail in a future publication (Cooper et al. in prep.). The specific characteristics of the COCO simulations are listed in Table  \ref{tab:cosmology}.

%\begin{itemize}
%\item High resolution volume \(L \sim 17.4 h^{−1} Mpc\)
%\item 
%High resolution volume 
%\apcedit{Radius of high resolution region} $L \sim 17.4\, h^{-1} \mathrm{Mpc}$
%\item DM particle mass \(1.135 × 10^{5} h^{−1} M_{\odot}\)
%\item DM particle mass $1.135 \times 10^5 h^{-1} \,\mathrm{M_{\odot}}$
%\item $10^{7}$ particles / halo
%\end{itemize}

%The COCO simulations are based on a cosmology model with the following parameters: 

%$\Omega_{\Lambda,0}=0.728, \quad \Omega_{m,0}=0.272, \quad \Omega_{b,0} = 0.04455, \quad \sigma_{8} = 0.81, \quad n_{S} = 0.967 \quad and \quad h = 0.704$.

\subsection{TNG50}
\label{sec:tng50}

\textit{IllustrisTNG} is a suite of large volume, cosmological, magnetohydrodynamical simulations run with the moving-mesh code AREPO \citep{springel2010} and with publicly available data \citep{nelson2019a}. One of such runs, the highest resolution, is known as \textit{TNG50-1}, or simply \textit{TNG50} \citep{pillepich2019, nelson2019b} and is used in this work. Its key characteristics are listed in Table \ref{tab:cosmology}.

%\begin{itemize}
%\item Volume \(L = 50 Mpc\)
%\item Baryonic mass \(8.5 \times 10^{4} M_{\odot}\) 
%\item DM mass, \(4.5 \times 10^{5} M_{\odot}\)
%\item $2160^3$ DM particles and $2160^3$ gas cells
%\end{itemize}

% The TNG50 simulations are based on a cosmology model consistent with the Planck 2015 results ($\Omega_{\Lambda,0}=0.6911, \quad \Omega_{m,0}=0.3089, \quad \Omega_{b,0} = 0.0486, \quad \sigma_{8} = 0.8159, \quad n_{S} = 0.9667 \quad and \quad h = 0.6774$).

The TNG50 simulation includes a comprehensive model for galaxy formation physics, which is able to realistically follow the formation and evolution of galaxies across cosmic time \citep{weinberger2017,pillepich2018}. TNG50 self-consistently solves the coupled evolution of dark matter, cosmic gas, luminous stars, and supermassive black-holes from a starting redshift of $z=127$ to the present day, $z=0$. In this work, we focus on the snapshots at $z < 0.02$.

\subsection{Auriga}
\label{sec:auriga}
 
The \textit{Auriga simulations} \citep{grand2017,grand2024} are a set of cosmological zoom-in magnetohydrodynamical simulations
%based on N-body magneto-%hydrodynamics moving mesh %code AREPO (Springel 2010). 
also carried out with the AREPO code.
Auriga
%imulations are based on dark matter halos selected by the 
re-simulates at higher resolution a sample of halos selected by the
EAGLE project \citep{schaye2015} and implements models for black hole (BH) accretion and feedback, stellar feedback, stellar evolution, chemical evolution, metallicity dependent cooling, star formation and magnetic fields, as reported in \citet{grand2017}.
%and implement a full baryonic physics modelling.
From the available simulations in the Auriga portal\footnote{\url{https://wwwmpa.mpa-garching.mpg.de/auriga/data.html}} we %have selected
use all 30 halos in the Original/4 series.

%with the following characteristics:

%\begin{itemize}

%\item 30 halos of \(1 < M_{200} / 10^{12}\, \mathrm{M_{\odot}} < 2\)
%\item Cube of \(L = 100\, h^{-1}\,\mathrm{Mpc}\) 
%\item baryonic mass \(~ 5 \times 10^{4}\,\mathrm{M_{\odot}}\)
%\item DM mass \(~ 3 \times 10^{5}\,\mathrm{M_{\odot}}\) 

%\end{itemize}

%The simulations are based on the $\Lambda$CDM cosmology, modelled with the following parameters: \(\Omega_{m} = 0.307,\quad \Omega_{\textit{b}}  = 0.048, \quad \Omega_{\Lambda} = 0.693\), and Hubble’s constant \(H_{0} = 100 \quad h km s^{-1} Mpc^{-1}\), $h = 0.6777$ \citep{planck2014}.

%The gravitational softening length for stellar and dark matter particles grows with scale factor up to a maximum of 369 pc. For the gas cells, the softening length scales with the mean radius of the cell but is not allowed to drop below the stellar softening length.

\begin{figure}[h!]
\centering
\includegraphics[width=0.85\columnwidth]{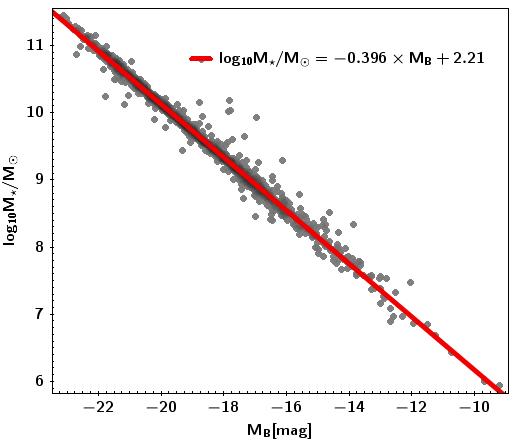}
  \caption{Stellar mass versus B-band absolute magnitude from the Spitzer S$^4$G catalogue. 
  The red line represents the empirical correlation
  in equation~\ref{eq:mstar_mb_relation}.}
  %$log_{10} M_{\star}/M_{\odot} = -0.396 \times M_{B} + 2.21$} 
  \label{fig-stellarmassvsmb}
\end{figure}

\begin{figure}[h!]
\centering
\includegraphics[width=0.8\columnwidth]{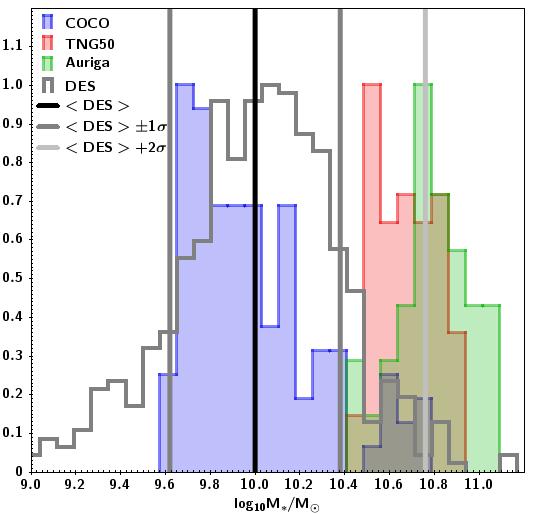}
  \caption{Histogram of 
  %the 
  $\log_{10} {M_\star}$
  for the observed (grey) and simulated galaxy samples (blue, red and green for COCO, TNG50 and Auriga, respectively).
  %of the stellar mass %distributions. 
  %\textit{grey outline}: stellar mass distribution of DES sample. 
  %\textit{blue}: stellar mass distribution of the COCO halos. \textit{red}: stellar mass distribution of the TNG50 halos. \textit{green}: stellar mass distribution of the Auriga halos. 
  The \textit{black vertical line} shows the average of the DES sample ($\log_{10} M_{\star} / \mathrm{M_{\odot}} = 10.00$) and the \textit{dark grey vertical lines} the
  %: average of the DES sampler 
  $\pm$ 1$\sigma$ range around that average, [$9.62, 10.38 $] . The \textit{light grey vertical line} shows the average of the DES sample $+2\sigma = 10.76$. 
  %Due to the different size of the populations, to increase readability, the maxima of all the 
  The maximum value of each histogram has been re-scaled to 1 for this comparison}.
  %have been rescaled to distributions is normalized to 1.}
  \label{fig-stellarmassCOCODES}
\end{figure}

\begin{figure}[h!]
\centering
\includegraphics[width=0.8\columnwidth]{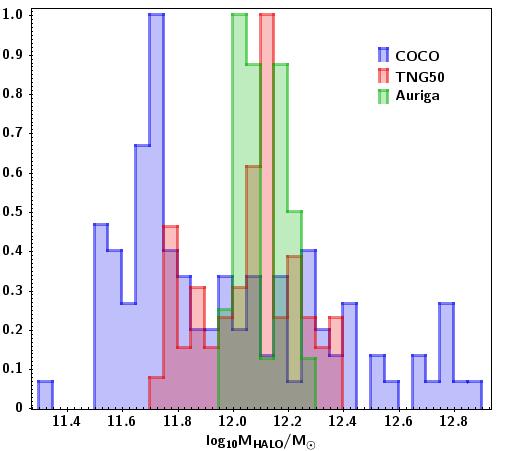}
  \caption{Histogram of $\log_{10}$ halo mass. Colors correspond to simulation sets as shown in the legend and Figure ~\ref{fig-stellarmassCOCODES}.
    The maximum value of each histogram has been re-scaled to 1 for this comparison}. %distributions. \textit{blue}: halo mass distribution of the COCO halos. \textit{red}: halo mass distribution of the TNG50 halos. \textit{green}: halo mass distribution of the Auriga halos. Due to the different size of the populations, to increase readability, the maxima of all the distributions is normalized to 1.}
  \label{fig-halomass}
\end{figure}

\begin{figure}[h!]
\centering
\includegraphics[width=1.0\columnwidth]{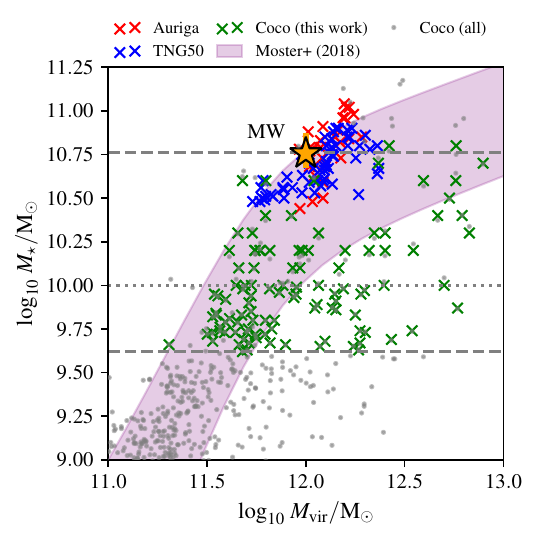}
  \caption{Stellar mass vs halo mass for the selected sample of galaxies in the COCO, TNG50 and Auriga simulations compared with the empirical correlation by \citet{moster2018} $\pm$ $2\sigma$. The horizontal lines indicate the DES galaxy sample stellar mass distribution: mean value (dotted line); mean value $- 1\sigma$ (lower dashed line) and mean value $+ 2\sigma$ (upper dashed line). The yellow star represents the MW.}
  \label{fig-smhm}
\end{figure}

\section{Halo Selection}
\label{sec:haloselection} 

In order to allow for a consistent comparison between the cosmological simulations and the DES image sample, we selected samples of simulated halos in a comparable range of both stellar mass and halo mass.
%a selection of the simulated halos has been made on the basis of their halo mass and stellar mass.  
To determine the range of host stellar mass in the DES sample, we %obtained the host stellar masses for the DES galaxy sample from 
construct an empirical relation between the absolute magnitude of the 
%sample host 
galaxies in the B-band, $M_{B}$, available from the HyperLeda database\footnote{\url{http://leda.univ-lyon1.fr/}} \citep{makarov2014}, and 
%the  
%. To this effect, as part of this work work, we have derived a correlation between (M$_{B}$) and 
%stellar mass by approximating the stellar mass vs M$_{B}$ relation for the galaxies observed in
their stellar mass, $M_\star$, as determined by
the Spitzer Survey of Stellar Structure in Galaxies \citep[S$^4$G,][]{sheth2010}. %The empirical correlation resulting from this approximation can be expressed, with a certainty of 0.995, by the following equation:   
We find the following relation:
\begin{equation}
\label{eq:mstar_mb_relation}
\log_{10} (M_* / \mathrm{M_{\odot}}) = - 0.396 \,M_{B} + 2.21
\end{equation}

%log (M$^*$ / M$_{\odot}$) = $- 0.396$ x M$_{B}$ + $2.21$ 

%$\log_{10} M_\star/\mathrm{M_\odot}$ = $- 0.396$ x M$_{B}$ + $2.21$

As shown in Figure~\ref{fig-stellarmassvsmb}, this empirical relation is linear over the magnitude range of the DES host sample, with small scatter (generally less than $0.2 \log_{10} (M_* / \mathrm{M_{\odot}})$ bar a few outliers).
%
%This empirical correlation 
%yields a fit of the data with a low level of scatter %as can be seen in Figure \ref{fig-stellarmassvsmb}.

The histogram in Figure \ref{fig-stellarmassCOCODES} shows the stellar mass distribution of the DES sample and of the selected halos in the COCO, TNG50 and Auriga simulations. 
The average stellar mass of the DES sample is $\log_{10} M_\star/\mathrm{M_\odot}$ = 10.00 and the standard deviation $\sigma = 0.38$.
While the average of the distributions is not the same, there is sufficient overlap between the DES sample stellar mass range and the stellar mass range of the simulated halos selected for the comparison.
Figure \ref{fig-halomass} shows the halo mass distribution of the selected halos in the COCO, TNG50 and Auriga simulations. Here all the simulations overlap and the average values are close to one another.
% (the maxima of the histograms is normalised to 1).
In this stellar mass range, COCO galaxies occupy a broader range of halo masses than TNG50 galaxies.

Figure \ref{fig-smhm} shows the stellar mass versus the halo mass for the selected sample of galaxies in the COCO, TNG50 and Auriga simulations compared with the empirical correlation by \citet{moster2018} $\pm$ $2\sigma$. The horizontal lines indicate the mean value, mean value $- 1\sigma$ and mean value $+ 2\sigma$ for the DES galaxy sample stellar mass distribution. While TNG50 galaxies are within Moster's correlation $\pm$ $2\sigma$, some Auriga galaxies seem to be slightly above the $= 2\sigma$ correlation range. COCO shows a larger scatter of stellar mass in the region above $\log_{10} M_{halo}/M_{\odot}$ and has a rather sparse sample of galaxies around the MW region. These COCO simulation characteristics are discussed in depth in Cooper et al. in prep.
%For TNG50 and Auriga, the simulated halos have a stellar mass corresponding to MW-like galaxies, see Figure \ref{fig-stellarmassCOCODES}.

\section{Mock Images}
\label{sec:mockimages}

%We aim 
To compare the predictions of cosmological simulations with the observations from the DES sample,
%, and for this purpose 
we have generated mock images from
snapshots of the simulations described in Section~\ref{sec:cosmosimulations} in the redshift range $0 < z < 0.02$. 
%the results of the the simulations. In this work we have used snapshots from the COCO, TNG50 and Auriga cosmological simulations for redshift $0 < z < 0.02$ as the predictive references (see Section \ref{sec:cosmosimulations} for an overview of these simulations).
The continuous stellar mass density field in all these simulations is represented by discrete tracers, called \textit{star particles}\footnote{In the COCO simulation, stellar mass is associated with a subset of the collisionless particles in post-processing, rather than an independent particle species in the original simulation, but the principle is the same from the point of view of our analysis; more details are given in \citet{cooper2017}.}. The stellar mass associated with each star particle corresponds to a stellar population with a single age and metallicity. As in any $N$-body realization of a density field, each particle notionally corresponds to an irregular volume of phase space centred on the location of the particle.
%
% The snapshots contain a 3D field of the dark matters and stellar particles mass distributions, and particles or a grid representing the gaseous component. The simulation also provides many properties for the stars gas such as ages and metallicity among many others. Here we are only interested in the stellar mass distribution and on the age and metallicity properties of the stellar particles.
%
%The snapshots contain a 3D field of particles, of a certain mass, age and metallicity in the halo of the host galaxy, including in-situ and accreted mass.
We have applied 
%a number of 
the following
transformations 
to the properties of the star particles from the simulation snapshots in order to recover the observables that can be identified in real images: %The generation of mock images from this output entails the following processing steps:

\begin{itemize}
\item Expansion of the discrete star particles into an approximation of the implied continuous 3-dimensional of stellar mass distribution, by convolution with an adaptive smoothing kernel;
\item Projection of the continuous distribution of stellar mass into a 2-dimensional plane. The orientation of the central galaxy relative to the observer's line of sight is a parameter of our method, and can be either random or specific (for example, to view the galaxy face-on or edge-on);
%\item Projection of 3D particle field (output of cosmological simulations) on a 2D plane, whose orientation can be selected by setting the unit vectors along the axis x,y,z. This allows to have a random orientation (such as the obtained observational surveys) or select a specific orientation such as face-on or edge-on. 
\item Conversion of stellar mass density to luminosity density, by convolution of an SED appropriate to the age and metallicity of each particle with a specific photometric bandpass.
%convolution with the photometric filter(s) used.
%\item Smoothing of the luminosity (distribution of luminosity on each pixel)
%\item Depending on the purpose, the application of a specific (instrument) PSF might be required.
\end{itemize}

\begin{figure}[h!]
\centering
\includegraphics[width=0.9\columnwidth]{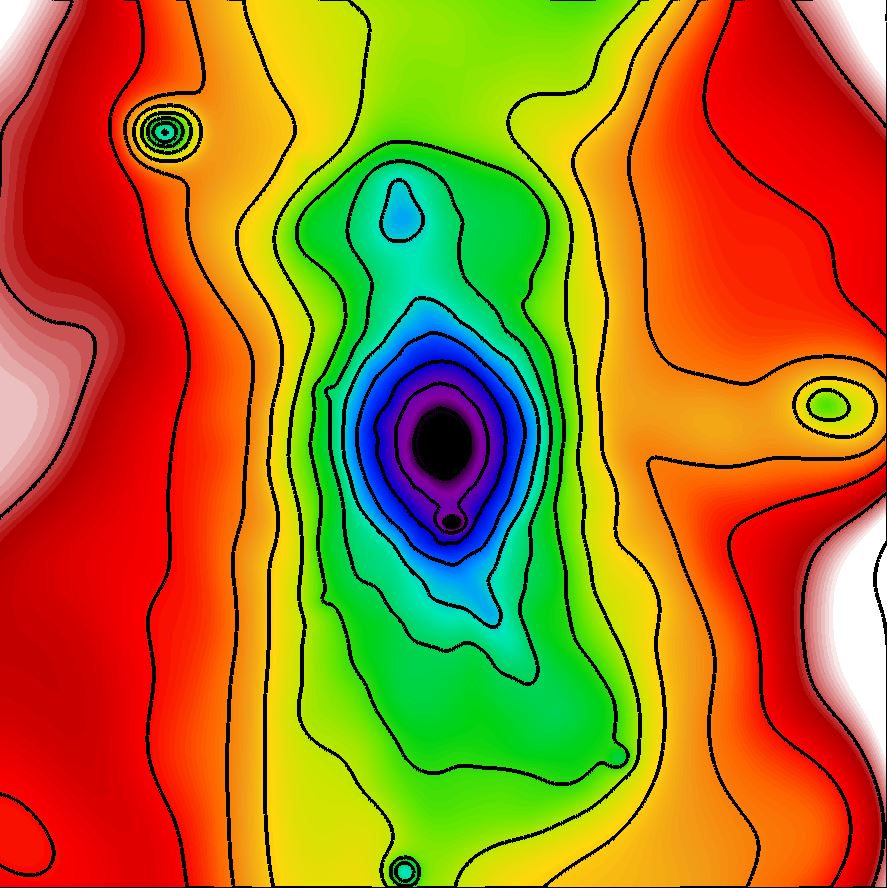}
  \caption{Example of a surface brightness map produced with \textit{pNbody}.
  %output of pNbody. 
  The \textit{black} contours mark the isophotes of surface brightness, starting with 20 mag arcsec$^{-2}$ in the centre of the host galaxy and separated by increments of 1 mag arcsec$^{-2}$.} 
  \label{fig-sbmap}
\end{figure}

%We have selected 
We use
the open source tool \textit{pNbody} \footnote{\url{https://obswww.unige.ch/~revaz/pNbody/}} \citep{revaz2013} to produce mock images by implementing the transformations above.
%the above steps.
%
%The input of \textit{pNbody} is a \textit{.hdf5} 
%file containing the snapshot of the cosmological simulation. The output of \textit{pNbody} is a 2D surface brightness map. 
Figure~\ref{fig-sbmap} shows, as an example, the $r$-band surface brightness map produced by processing one of the COCO galaxies with \textit{pNbody}. 
%\textit{pNbody} for the $r$-band obtained from one of the COCO halos. 
The contour lines identify the isophotes for an intuitive view of the possible low surface brightness (LSB) structures present in the image. 
%We then transform the surface brightness map to an image with counts. 

We have carried out tests to asses the 
robustness 
%sensitivity 
of the resulting mock-images 
%from 
to variations in
the \textit{pNbody} configuration parameters for the generation of %realistic 
mock images with the 
DECam instrument. 
We have generated surface brightness maps for a halo at a distance of 70 Mpc from the Sun, as a representative distance for the DES galaxy sample that ranges from 40 to 100 Mpc (changing the distance within the DES galaxy sample range does not noticeably influence the detectability of streams, as explained in Section  \ref{sec:streamdetection}). 
%We have varied selected input parameters, notably the smoothing length, the image size, pixel scale and distance to the source and compared the results, in particular the resulting surface brightness measured in apertures of equivalent size in kpc. Also the surface brightness radial profile of the central galaxy has been compared between the observations and simulations for selected images. 

For the expansion (smoothing) step, we distribute the luminosity (flux) of each particle over the image pixels by convolution with a Gaussian kernel. The scale of this kernel, $h$, which we refer to as the \textit{smoothing scale}, is set to a parametrized multiple to the root-mean-square average distance to the 16$^{th}$-nearest star particle neighbour, $h_{16}$. It therefore adapts to the local star particle density. The logic for setting $h$ is broadly similar to that used to determine the kernel scale in a smoothed particle hydrodynamics calculation. However, the smoothing in our case only serves to interpolate between the original particles and does not have any physical significance. It is therefore somewhat arbitrary; our method reflects a balance between smoothing sufficiently to reduce the visual impression of a discrete particle distribution, while preserving the small-scale features.

\begin{figure*}[h!]
\centering
\includegraphics[width=1.0\textwidth]{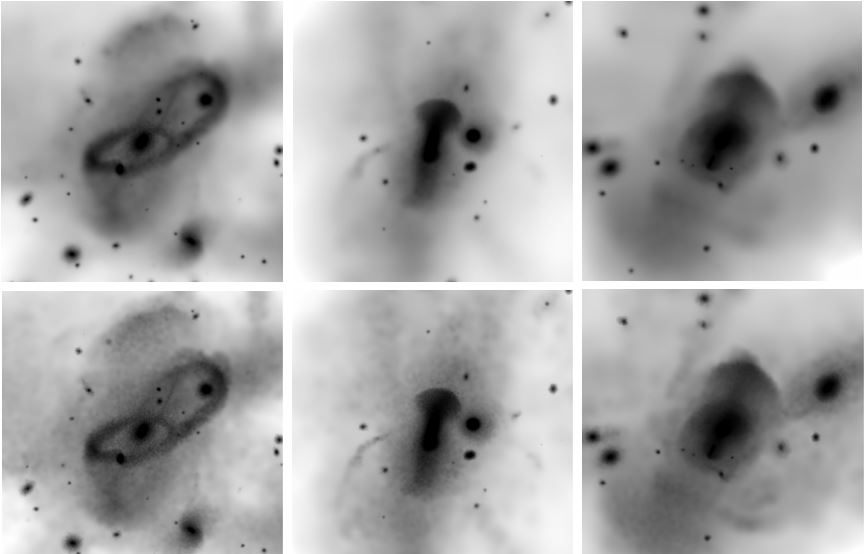}
  \caption{Comparison of smoothing methods. The top panels show the the surface brightness map of three different halos generated with a smoothing length up to the 16th neighbour multiplied by the factor 0.6. The bottom panels show the surface brightness maps of the same halos but applying a factor of $0.3$ to the smoothing length to the 16th neighbour.} 
  \label{fig-smoothing2}
\end{figure*}

% A key step in the process of generating mock  images from cosmological simulations is the distribution of the luminosity of the particles (and thereby the flux) across the image pixels. This is done by defining the so called \textit{smoothing length}. In order to analyse the influence of the choice of smoothing length on the detectability of streams in the mock-images, several implementations of smoothing were tested and their impact on the resulting surface brightness map analysed. 
%\ref{fig-smoothing1} and 
Figure \ref{fig-smoothing2} shows 
%the result of applying a 
%smoothing length equal to the distance to the 16th nearest neighbour (right panel), and somewhat reduced smoothing lengths, by a factor of $0.3$ (left panel) and $0.6$ (middle panel) of the initial smoothing length.
mock images of the same three galaxies (left to right) with kernels of scale $h=0.6\,h_{16}$ (top row) and $h=0.3\,h_{16}$ (bottom row). Over this range (with the DESI pixel scale) the bulk of the particle granularity in the image is removed, but nature, extent and surface brightness of tidal features relevant to our subsequent visual inspection (see Section \ref{sec:streamdetection}) are not noticeably different. 
%The results indicate that these changes in the smoothing length do not modify noticeably the perception of the surface brightness map, important for the subsequent visual inspection exercise (see Section \ref{sec:streamdetection}).

%\begin{figure*}[h!]
%\centering
%\includegraphics[width=1.0\textwidth]{Smoothing-length-factor-compariron.jpg}
%  \caption{Comparison of smoothing methods. The right panel shows the the surface brightness map of a halo with an smoothing length up to the 16th neighbour. Left and middle panels show the result of applying a factor of $0.6$ (middle) and $0.3$ (left) to the smoothing length as calculated to the 16th neighbour.} 
%  \label{fig-smoothing1}
%\end{figure*}

\begin{figure*}[h!]
\centering
\includegraphics[width=1.0\textwidth]{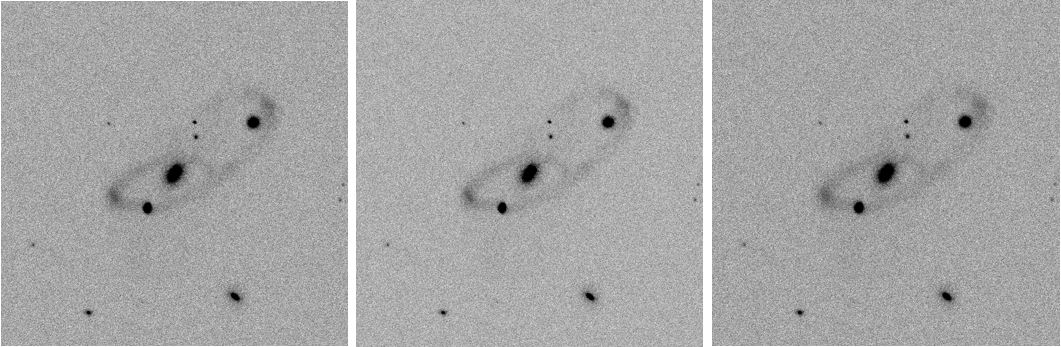}
  \caption{Comparison of smoothing methods. Images generated from the surface brightness maps, by transforming the surface brightness to counts, and adding background noise to emulate a surface brightness limit of 29 mag arcsec$^{-2}$. Left and centre images correspond to a smoothing length equal to the distances to the 16th nearest neighbour multiplied by a factor of $0.3$ (left) and a factor of $0.6$ (centre).  The right panel shows the image produced with a different smoothing length, considering the distance to the 5th neighbour.} 
  \label{fig-smoothing3}
\end{figure*}

We use the mock images i) to assess the detectability of streams over a range of surface brightness limits and ii) to measure the photometry of the streams that would be detectable at the surface brightness limit of the DES observations. 
To follow the same process of stellar stream detection by visual inspection as with real images, we have transformed the surface brightness maps into counts images. Then we have added background noise to the images, which depends on the type of analysis to be carried out.

In order to assess the detectability of the streams under different image depths, a simple realization of background noise was added to the images, as flat Gaussian noise with variable amplitude. This gives us the flexibility to emulate different surface brightness limits in an easy way; we choose limits between 25 and 34 mag arcsec$^{-2}$ in intervals of 1 mag arcsec$^{-2}$. %This has been carried out with 
For this purpose we used the state-of-the-art  {\it GNU Astronomy Utilities} (Gnuastro)\footnote{\url{http://www.gnu.org/software/gnuastro}} software.

To asses the impact of the smoothing length on the detection of a possible stream in the mock images by visual inspection, we have examined count images with different choices of smoothing length, after adding background noise corresponding to a surface brightness limit of 29 mag arcsec$^{-2}$.
Figure~\ref{fig-smoothing3} shows this test for smoothing lengths $h=0.6\,h_{16}$, $h=0.3\,h_{16}$ and an alternative smoothing scheme in which $h$ is instead set to the \textit{absolute} distance to the $5^{th}$ nearest neighbour.
%No noticeable difference can be appreciated through visual inspection of these images.
Again, regarding stream detection by visual inspection, we find no significant difference between the images.
We therefore adopt a smoothing scale of $h=0.6\,h_{16}$.
%It is therefore concluded that the smoothing methods applied here do not have a noticeable impact on the ability to detect streams by visual inspection. Thus, we have selected a smoothing length equal to $0.6$ times the distance to the 16th nearest neighbour.

To measure photometric properties of the mock images and compare the results with those of the DES sample, we add a more realistic sky background to the mock images. We first extract the sky background from selected real DES sample images, 
removing the central galaxy and replacing it by real sky background from an area of the same image without significant point sources.
The image from which we extracted this fiducial background image was selected according to the following criteria: i) the central galaxy should be edge-on, in order to minimise its area of influence on the image; ii) the image should have no stream detection, in order not to interfere with the synthetic halo image; and iii) the image should have an $r$-band surface brightness limit representative of the DES sample.  
The selected image had a surface brightness limit of 28.65 mag arcsec$^{-2}$.  
The fiducial sky background extracted from this image, as above, was then superimposed onto the mock images generated directly from the surface brightness maps, that have only a central galaxy and possibly other accompanying galaxies but no sky background, creating an image with a synthetic central galaxy and a real DES background. 
 
We use \textit{Gnuastro} to replace the central galaxy and its surroundings with a section of the same image  extracted from a region without bright sources. The size and orientation of the ellipsoidal mask is such that covers the area of influence of the central galaxy in its surroundings, i.e. it reaches up to the point where the surface brightness profile flattens.

%A schematic depiction of this process can be seen in Figure \ref{fig-skyaddition-1}. Figure \ref{fig-skyaddition-2} presents a block diagram of the actions carried out out to obtain an image with 'only' sky background with the Gnuastro tool.

\begin{figure*}[h!]
\centering
\includegraphics[width=1.0\textwidth]{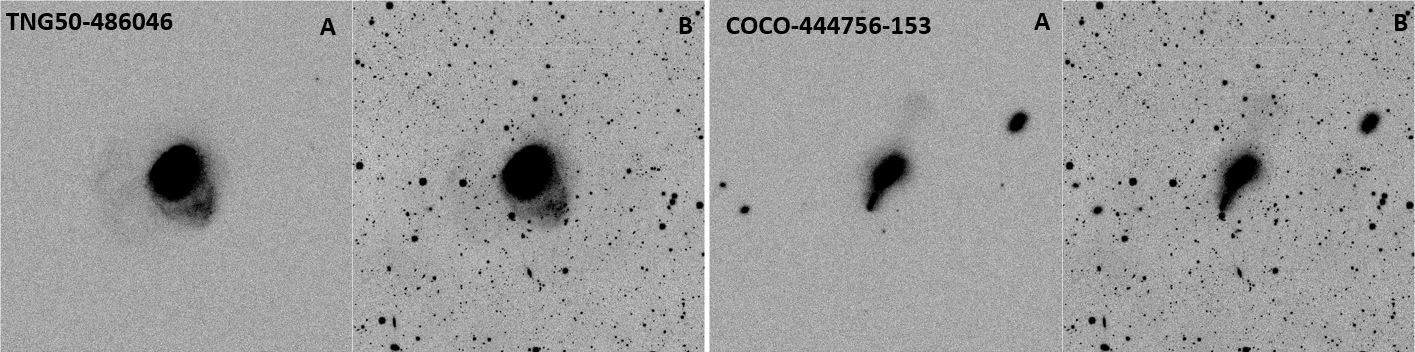}
  \caption{Comparison of mock images with simulated and real DES sky background: examples of a TNG50 (left) and a COCO (right) simulated galaxies with streams stacked with a real DES sky background (panels B), and with artificial flat background (panels A) both having the same surface brightness limit of 28.65 [mag arcsec$^{-2}$].} 
  \label{fig-skynoisebackground}
\end{figure*}

Figure~\ref{fig-skynoisebackground} shows examples of TNG50 and COCO mock images with streams around the central galaxies. Panels B show images to which we have added the sky background of a real DES image with a surface brightness limit of $28.65\,\mathrm{mag\,arcsec^{-2}}$. We compare these images to those in panels A, to which we have added an artificial Gaussian sky background equivalent to the same surface brightness limit.
%with artificial background corresponding to the same surface brightness limit used in Section \ref{sec:streamdetection} for analysing the detectability of streams under different image depths. 
The streams, though faint, can be clearly appreciated in the images with real DES background and, although they are more obvious in the image with artificial background, the their appearance is consistent in both images. 
Consequently, adding a synthetic background to the mock images does not seem to impact significantly the detection of streams by means of visual inspection with respect to the real background in this surface brightness limit regime. This method is very efficient to emulate different levels of surface brightness limit and we adopt it in this work to assess the detectability of streams at those levels currently not achieved in available surveys. However we acknowledge that for much fainter surface brightness limits, this method provides only an approximation as the confusion of sources and possibly cirri will become much more significant and the difference between the synthetic background and the real background will increase.

\section{Stream Detection}
\label{sec:streamdetection}

In this section, the detectability of tidal streams under different image depths is analysed on the basis of mock-images. At the present time, there is no automatic method for detecting tidal streams, though this is an important subject of current research.
%(see Section \ref{sec:futureresearch} for a review of current work in this area). 
Therefore detectability is based here on the method combining visual inspection and image analysis tools, as presented in Section \ref{sec:cosmosimulations}. This is the method applied to the detection of streams in real images, as presented in \citet{miro-carretero2024} and that has been the basis for generating the stream catalogue presented there.

\begin{comment}
We made all the photometry measurements by applying Gnuastro's {\sc MakeCatalog} subroutine \citep{akhlaghi2019a} on the sky-subtracted images generated by Gnuastro's {\sc NoiseChisel} \citep{akhlaghi2015,akhlaghi2019b}. The method is identical to the one applied to the analysis of the DES sample images and explained in detail in \citet{miro-carretero2024}. 
\end{comment}

The analysis has been done on the basis of the r-band images. This band has been chosen as it 
%has been the band for which the detection area was, with few exceptions, the largest when carrying out the photometry analysis with Gnuastro. Also, this band 
has been used in other relevant observational studies on tidal features \citep{miskolczi2011,morales2018,bilek2020,sola2022,skryabina2024}, thus allowing for comparison of results across different studies. In the mock-images, the SDSS $r$-filter produces brighter measurements of the stream than the SDSS $g$-filter, in agreement with the observations. Depending on the region in the image this difference can be up to $\sim$ 0.6 mag arcsec$^{-2}$.

In order to assess the impact of the host distance on the detectability of streams, mock images have been produced at different distances (50, 70 and 90 Mpc), covering the distance range of the observed hosts with their streams, which ranges from 40 to 100 Mpc. Although the surface brightness is independent of the distance, the distance has an impact on the S/N at pixel level that may in turn impact the observability. However, as expected, this  small relative difference in the distance does not show a noticeable difference in the visual perception of the images.
%, as can be observed in Figure \ref{fig-distancecomparison3}. 

%\begin{figure*}[h!]
%\centering
%\includegraphics[width=1.0\textwidth]{fig-distance-comparison.jpg}
%  \caption{Comparison of the effect of host distance on visual perception. Images generated from the surface brightness maps, by transforming the surface brightness to counts, and adding background noise to emulate a surface brightness limit of 29 mag arcsec$^{-2}$. The centre image (reference image) corresponds to a distance of 70 Mpc from the Sun to the host. In the left image the distance is 50 Mpc (image zoomed-out to match the size of the stream in the reference image). In the left image the distance is 90 Mpc (image zoomed-in to match the size of the stream in the reference image).} 
%  \label{fig-distancecomparison3}
%\end{figure*}

\begin{figure}[h!]
\centering
\includegraphics[width=0.8\columnwidth]{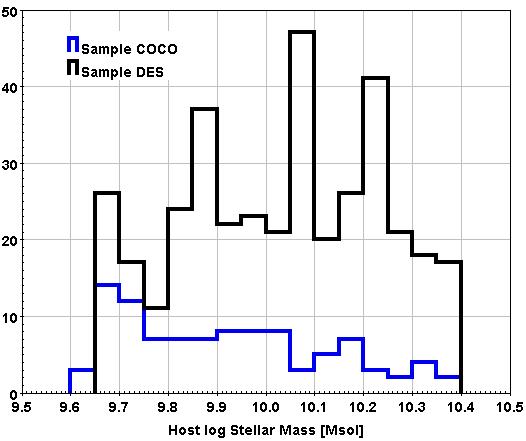}
  \caption{Distribution of host stellar mass in the DES sample and in the the COCO simulations within the selected stellar mass range.} 
  \label{fig-histogramstelllarmass}
\end{figure}

\begin{figure*}[h!]
\centering
\includegraphics[width=0.9\textwidth]{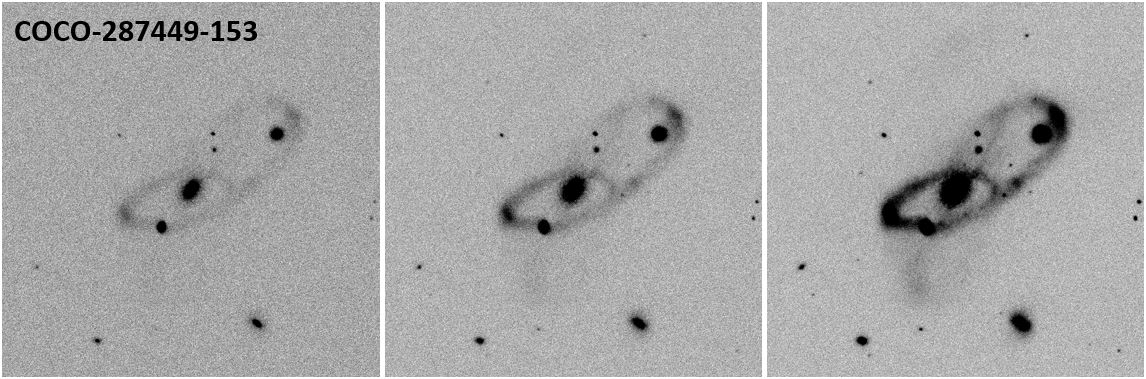}
\includegraphics[width=0.9\textwidth]{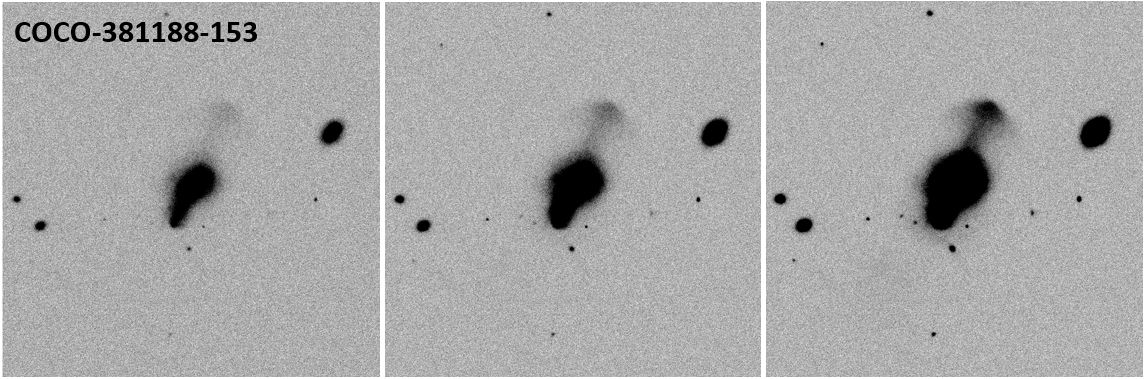}
\includegraphics[width=0.9\textwidth]{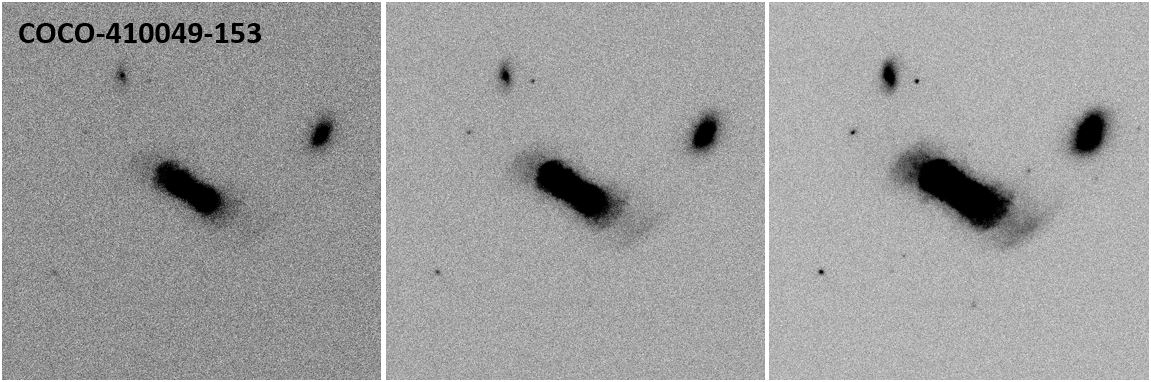}
\includegraphics[width=0.9\textwidth]{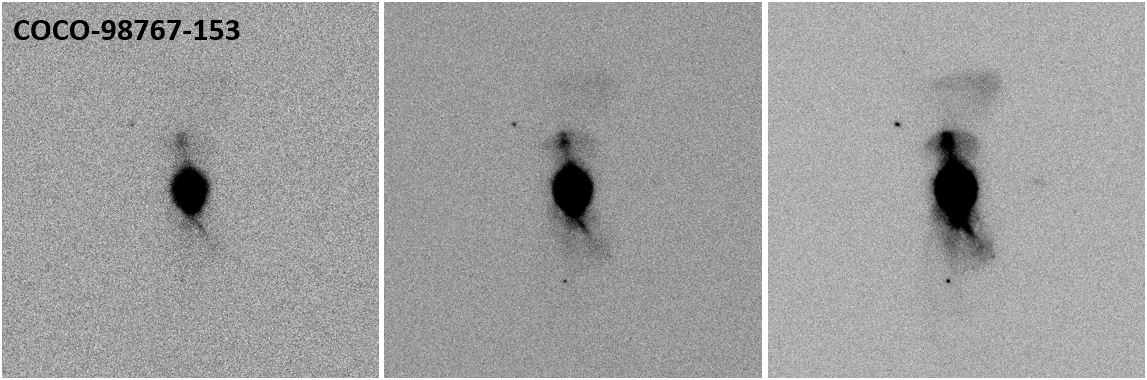}
  \caption{Examples of COCO mock-images COCO-287449-153, COCO-381188-153, COCO410049-153 and COCO-98767-153 at different surface brightness limits. From left to right 29, 30 and 31 mag arcsec$^{-2}$. The streams appear clearer as the depth of the image increases from left to right. All images are $10\times10$ arcmin.} 
  \label{fig-cocohaloexamples-1}
\end{figure*}

\begin{figure}[h!]
\centering
\includegraphics[width=0.9\columnwidth]{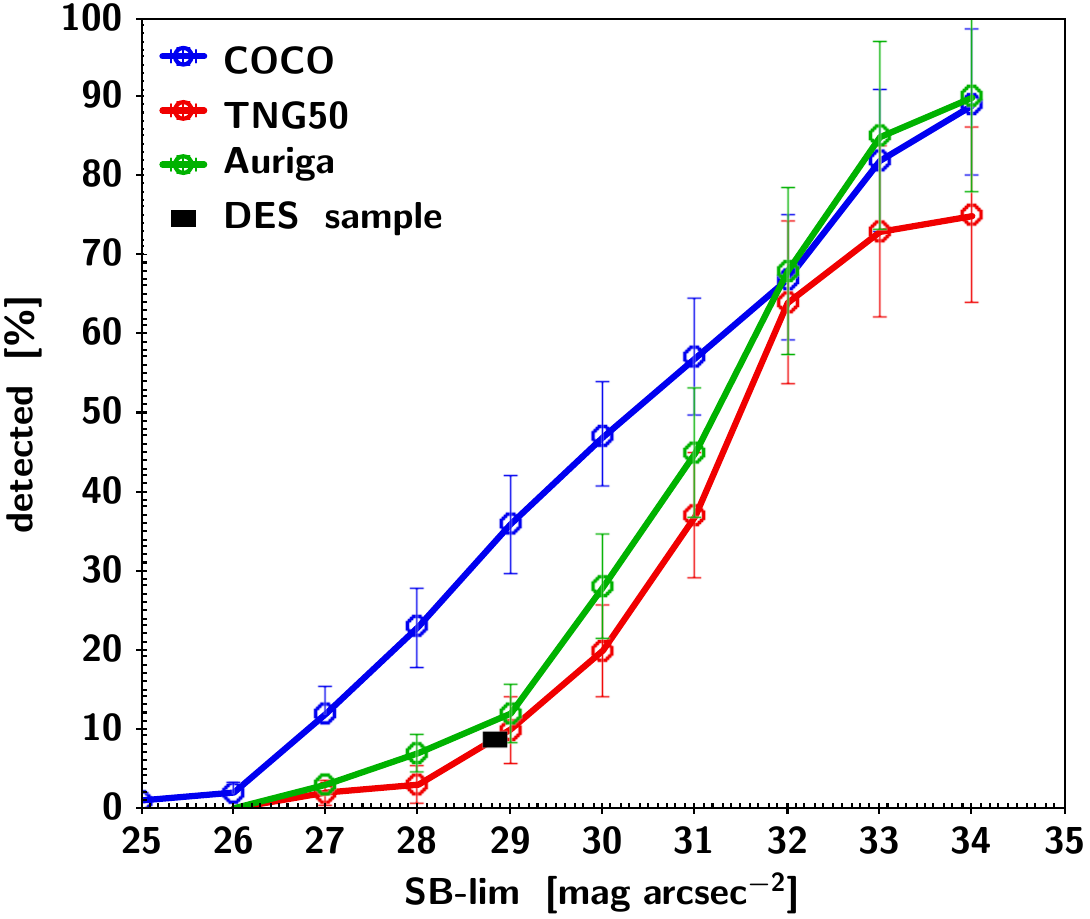}
  \caption{Detection rate curves for the COCO, TNG50 and Auriga cosmological simulations and the detection rate observed for the DES sample (\textit{black square} indicates the dispersion in surface brightness limit of the DES sample and the dispersion in the detection rate according to a binomial distribution, see \citet{miro-carretero2024}). The percentage of the streams detected by visual inspection is plotted versus the surface brightness limit at which such detection was possible. The bars indicate the Poisson confidence interval.}
  \label{fig-comparisondesonlyetectioncurve}
\end{figure}

%\begin{figure}[h!]
%\centering
%\includegraphics[width=0.8\columnwidth]{COCO_Stream-Detectability_240816.jpg}
%  \caption{Detection curve for the COCO simulations. The percentage of the streams detected by visual inspection is plotted versus the surface brightness limit at which such detection was possible. The bars indicate the Poisson confidence interval.} 
%  \label{fig-streamdetection}
%\end{figure}

\begin{figure}
\centering
\includegraphics[width=0.8\columnwidth]{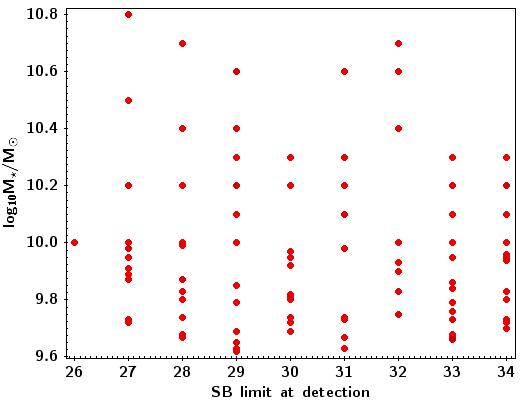}
  \caption{The plot shows the $log_{10} M_{\star} / M_{\odot}$ distribution of the COCO halos versus the surface brightness limit at which their streams are detected.} 
  \label{fig-cocomassstarvssblim}
\end{figure}

\subsection{Stellar Streams in the COCO Simulation}
\label{sec:cocosimulations}

For the comparison between the DES sample and the COCO simulations, in the first step we selected a sample of simulated halos containing central galaxies with
%halos of
stellar mass 
%within the range of 
in a range around
the DES stellar sample average $\pm$ 1$\sigma$, that is, between  $4.17 \times 10^9 M_{\odot}$ 
($\log_{10} M_\star/\mathrm{M_\odot}$ = 9.62) and $2.4 \times 10^{10} M_{\odot}$ 
($\log_{10} M_\star/\mathrm{M_\odot}$ = 10.38). 
In a second step, we selected a sample of simulated 
%halos
halos containing central galaxies with
stellar mass 
%within the range 
corresponding to the DES host sample average stellar mass value $+1\sigma$ and $+2\sigma$ ($\log_{10} M_\star/\mathrm{M_\odot} = 10.76$). 
%\apc{Does this mean the range $\mu+1\sigma < M_{\star} < \mu+2\sigma$ (where $\mu$ is the mean of the sample) or something else?}
%(log M/M$_{\odot}$ $\sim$ 7 for the COCO halos and $\sim$ 10 for the DES sample)

For the COCO simulations, 108 halos have been selected with central galaxy stellar masses overlapping with 70\% of the DES sample with streams.
A comparison of the host stellar mass distribution of the DES sample with the COCO simulation sample, within the selected stellar mass range, 
%can be seen 
is shown in Figure \ref{fig-histogramstelllarmass}

%The line of sight (an input parameter in the generation of surface brightness maps) was chosen in the $z$ coordinate direction (0,0,1) as in principle the orientation of the halos in the spacial coordinates is aleatory.

We have generated mock images with surface brightness limit  between 25 and 34 mag arcsec$^{-2}$ in intervals of 1 mag arcsec$^{-2}$. As the depth of the image increases, that is, the fainter the surface brightness limit is, the clearer the underlying streams appear to the observer carrying out the visual inspection and facilitating their detection, as can be seen in Figure \ref{fig-cocohaloexamples-1}.

%\begin{landscape}

%\begin{figure*}[h!]
%\centering
%\includegraphics[width=1.0\columnwidth]{COCO-images.jpg}
%  \caption{Examples of COCO mock-images at different surface brightness limits. At the top in blue the names of the halos are displayed. On the left in red the levels of surface brightness limit are shown, from shallower (top) to deeper (bottom). The streams appear clearer as the depth of the image from top to bottom.} 
%  \label{fig-cocohaloexamples-2}
%\end{figure*}

%\end{landscape}

Then the resulting $\sim 1000$ images were visually inspected and for each one, the surface brightness limit at which a streams could be visually detected for the first time was identified and noted. The inspection was carried out on the image   
FITS-format files, displayed with the SAO DS9 tool. As for the inspection of the observational images, suitable colour, scale and analysis block level options were selected to improve the perception. 
The DES observational images have a ‘real’ sky background while for the mock-images we have a sky background that while corresponding to a similar surface brightness limit, is much smoother, allowing for an easier detection of the streams. This has been taken into account in assessing the detectability of streams by reporting one level deeper when the stream was not clearly distinguishable by visual inspection.

The result is a curve indicating the percentage of streams detected within the COCO halo sample as a function of the image depth, that is, as a function of the image surface brightness limit. This is depicted in Figure \ref{fig-comparisondesonlyetectioncurve}. The curve follows an approximately linear trend between surface brightness limits of 26 and 34 mag  arcsec$^{-2}$, with a $12.5\%$ increase in streams detected per 1 mag arcsec$^{-2}$ increase in surface brightness limit. In $97\%$ of the mock-images, we detect by visual inspection what we consider to be (part of) a stream up to a SB-limit of 34 mag arcsec$^{-2}$.
However, we estimate that only 89\% of such faint structures could actually be measured with a reasonable level of error in real images.
   
Regarding the stream morphology, in the COCO sample, at SB-limit 28-29 arcsec$^{-2}$, around 70-80\% of the detected streams are shells --a segment of the wider morphology class known as umbrella, see stream morphology classification in \citet{miro-carretero2024}-- while at SB-limit 34 arcsec$^{-2}$ around 80-90\% of the detected streams are shells, with only 2-3\% of the streams displaying a circular morphology.

%A comparison of the stellar mass distribution for the galaxies with detected streams in the DES and CoCo simulation samples for the DES sample SB limit (28.26 [mag arcsec$^{-2}$]) and the corresponding 2 COCO simulations (28 and 29 [mag arcsec$^{-2}$]) can be seen in the following histogram

We have investigated a possible correlation between the surface brightness limit at which streams are detected and the host galaxy stellar mass. In particular, we have analysed whether more massive host galaxies show streams at a lower (brighter) surface brightness limit. Figure \ref{fig-cocomassstarvssblim} shows the stellar mass distribution for the galaxies identified as hosting streams versus the surface brightness limit at which those streams have been detected. 
%Based on that graph, such a 
No correlation is 
%not 
evident.

\begin{figure*}[h!]
\centering
\includegraphics[width=0.9\textwidth]{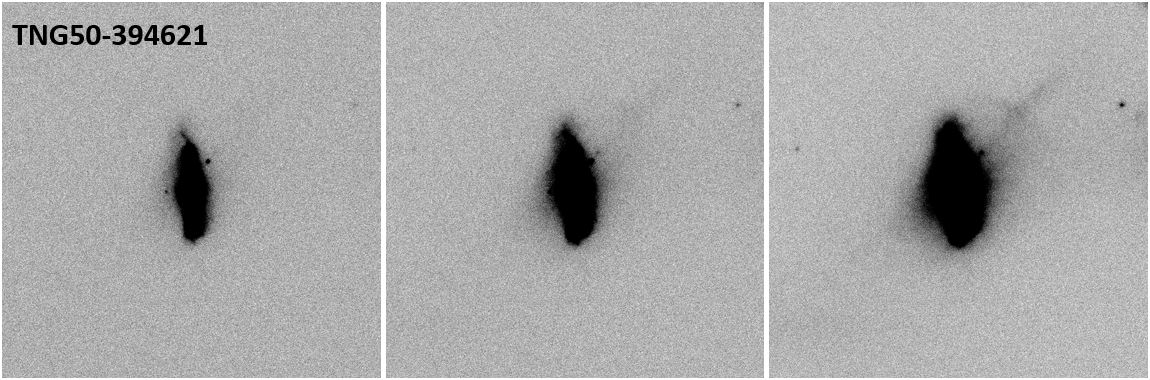}
\includegraphics[width=0.9\textwidth]{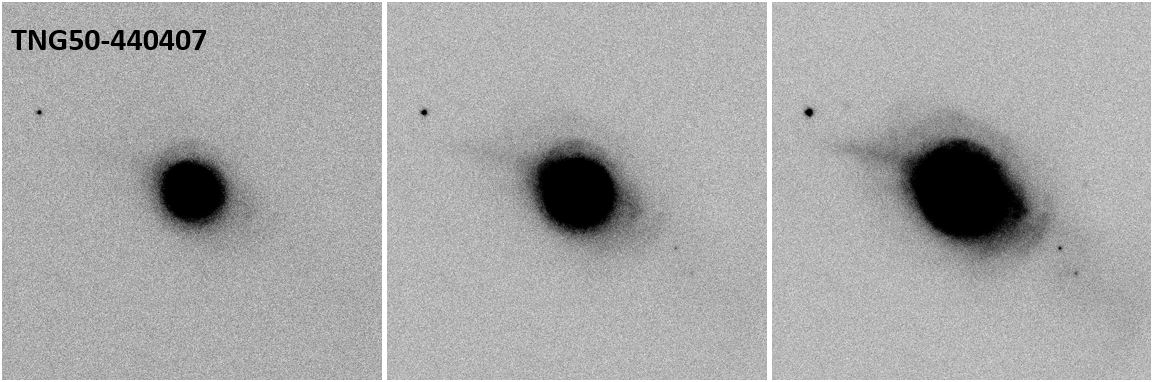}
\includegraphics[width=0.9\textwidth]{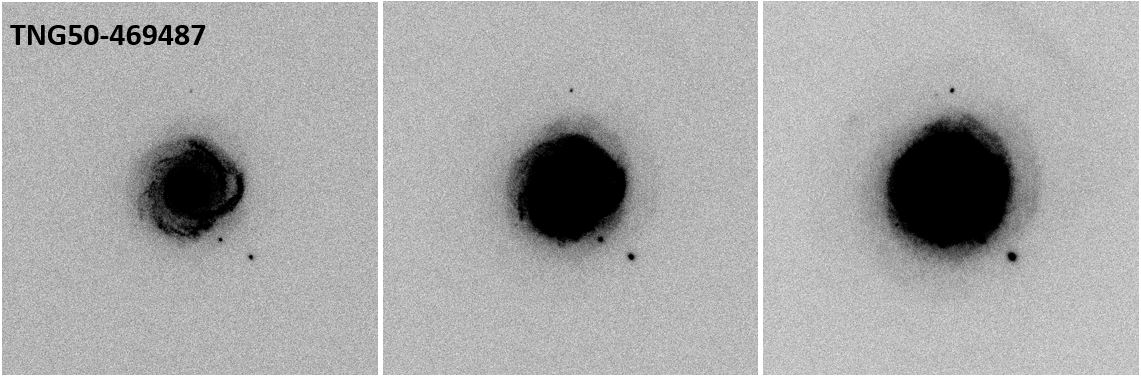}
\includegraphics[width=0.9\textwidth]{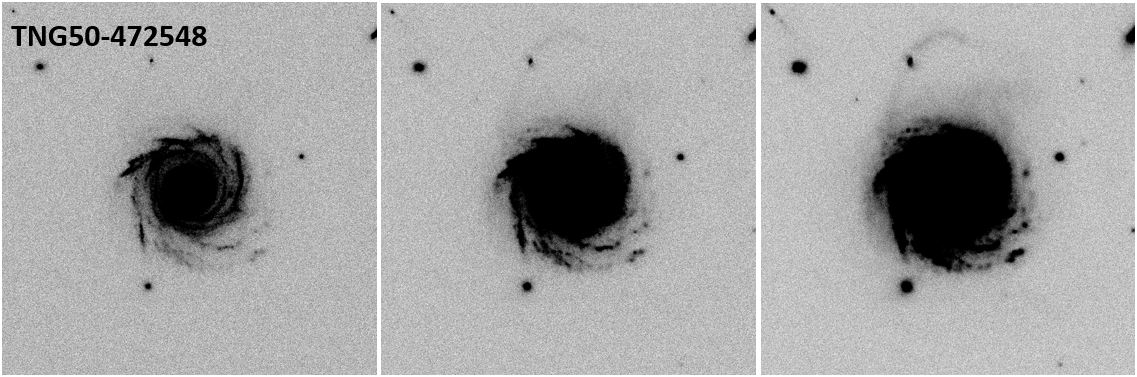}
  \caption{Examples of TNG50 mock-images TNG50-394621, TNG50-440407, TNG50-469487 and TNG50-472548 at different surface brightness limits. From left to right 29, 30 and 31 mag arcsec$^{-2}$. The streams appear clearer as the depth of the image increases from left to right. All images are $10\times10$ arcmin.} 
  \label{fig-tng50haloexamples-1}
\end{figure*}

\subsection{Stellar Streams in the TNG50 Simulation}
\label{sec:tng50simulations}

We have generated surface brightness maps from the TNG50 simulation for 60 halos.
40 of these have central galaxies in the stellar mass range between between  $3.02 \times 10^{10} M_{\odot}$ 
($\log_{10} M_\star/\mathrm{M_\odot}$ = 10.48) and $5.73 \times 10^{10} M_{\odot}$ 
($\log_{10} M_\star/\mathrm{M_\odot}$ = 10.76),
corresponding to the range of stellar mass between the average value of the DES sample $+$1$\sigma$ and $+$2$\sigma$. 
In order to compare with MW-like galaxies (such as those in the Auriga simulations) we have selected 20 additional halos in an extension of the stellar mass range to $8.0 \times 10^{10} M_{\odot}$ ( $\log_{10} M_{\star} / M_{\odot} = 10.9$), see Section \ref{sec:haloselection} for details.
 
We have transformed the surface brightness maps into images with counts and added sky background corresponding to 10 levels of surface brightness limit between 25 and 34 mag arcsec$^{-2}$ (see
Section \ref{sec:mockimages}).  As we increase the depth of the image (apply a lower background noise level to the image, corresponding to a fainter surface brightness limit), streams become more visible and can be detected by visual inspection, as can be seen in Figure  \ref{fig-tng50haloexamples-1}.
We have visually inspected the resulting $\sim 600$ images and for each halo / host galaxy, the surface brightness limit at which a streams could first be visually appreciated was identified and noted. 
%The inspection was carried out on the image \textit{.fits} files, displayed with the SAO ds9 tool. The result is a curve indicating the percentage of streams detected within the the TNG50 halo sample as a function of the image surface brightness limit.  
As with our analysis of COCO, FITS images were inspected with DS9.

%\begin{figure}
%\centering
%\includegraphics[width=1.0\columnwidth]{TNG50-detectability-curve_240328.jpg}
%  \caption{Detection curve for the TNG50 simulations. The percentage of the streams detected by visual inspection is plotted versus the surface brightness limit at which such detection was possible (blue curve). Where a faint structure was detected but could not be ascertained as a stream, the detection was nevertheless recorded (green carve).}
%  \label{fig-tng50streamdetection}
%\end{figure}

\begin{figure}[h!]
\centering
\includegraphics[width=0.9\columnwidth]{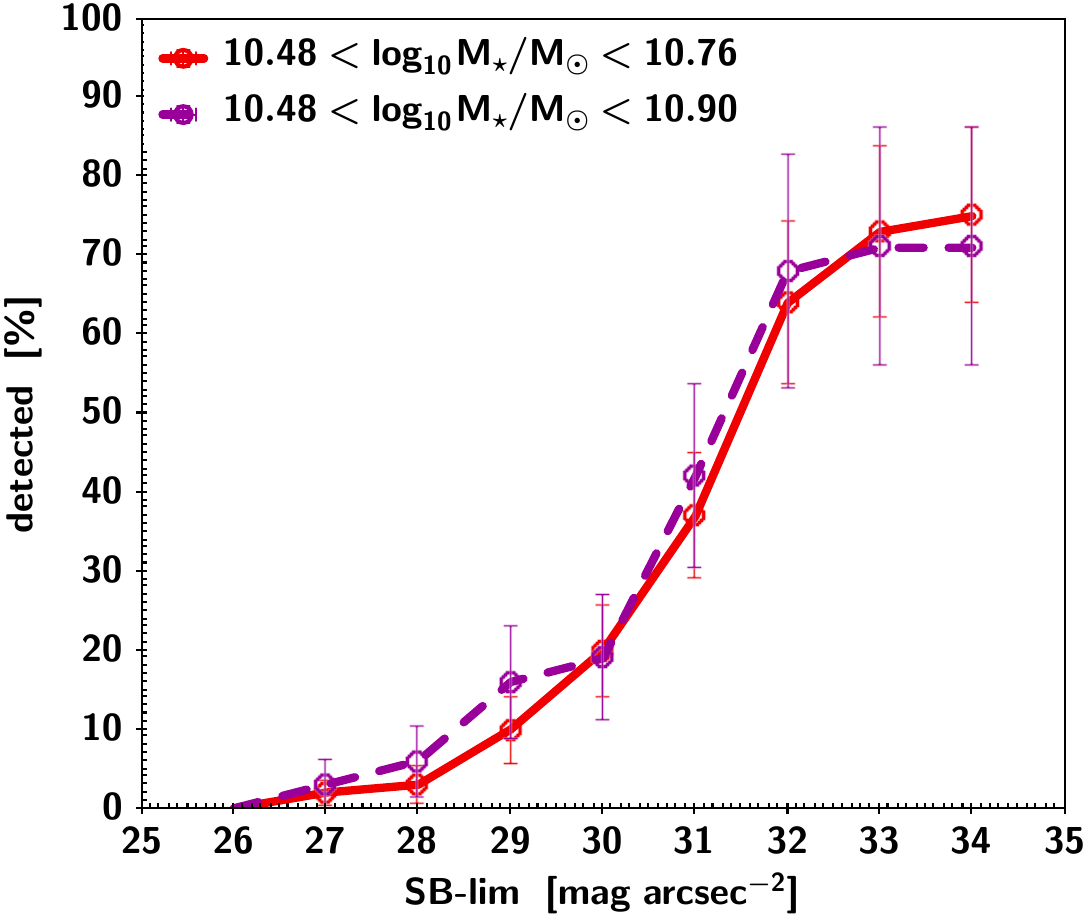}
  \caption{Stellar streams detection rate as a function of the image surface brightness limit for the TNG50 simulation. The percentage of the streams detected by visual inspection is plotted versus the surface brightness limit at which such detection was possible. The solid line corresponds to a stellar mass range of $\log_{10} M_\star/\mathrm{M_\odot}$ = 10.48  and  $\log_{10} M_\star/\mathrm{M_\odot}$ = 10.76, while the dashed line corresponds to an extended mass range up to $\log_{10} M_\star/\mathrm{M_\odot}$ = 10.9.}
  \label{fig-tng50detectionvsmass}
\end{figure}

%\begin{figure}
%\centering
%\includegraphics[width=1.0\columnwidth]{TNG50-COCO-detectability-curve_240328.jpg}
%  \caption{Comparison of the detection curves for TNG50 and COCO.}
%  \label{fig-tng50cocostreamdetection}
%\end{figure}

%The result is depicted in 

Figure \ref{fig-comparisondesonlyetectioncurve} shows the resulting curve indicating the percentage of streams detected within the the TNG50 halo sample as a function of the image surface brightness limit.
The curve shows a steep gradient between SB-limit = 30 and SB-limit=32 mag / arcsec$^2$, of about 25 $\%$ increase in streams detected per 1 mag  arcsec$^{-2}$ increase in surface brightness limit.
In $\sim$ 70\% of the mock-images a (part of a) stream can be detected by visual inspection at a SB-limit of 34 mag / arcsec$^2$.   
The detection percentage level resulting from inspection of the TNG50 mock-images at a SB-limit of 28.65 mag arcsec$^{-2}$ (the SB limit value corresponding to the average SB-limit for the DES sample in the $r$ band) is between 3\% and 10\% (corresponding to the SB-limit values for 28 and 29 mag arcsec$^{-2}$ in the curve, respectively). 

To assess the dependency of the detection rate for a certain SB-limit on the host stellar mass, we have compared the detectability curve obtained for 30 halos in the stellar mass range between $\log_{10} M_\star/\mathrm{M_\odot}$ = 10.48 and  $\log_{10} M_\star/\mathrm{M_\odot}$ = 10.76 with the one obtained for 60 halos in an extended stellar mass range up to $\log_{10} M_\star/\mathrm{M_\odot}$ = 10.9. As can be seen in Figure \ref{fig-tng50detectionvsmass}, both curves are very similar, the one for the extended mass range appearing smoother, due to the increased population of galaxies included. This seems to reinforce the speculation made in Section \ref{sec:cocosimulations} using COCO simulations that the stellar mass of the host galaxy does not noticeably influence the brightness of the streams around it. 

We have taken into account the fact that the smooth background added to the image makes %easier the detection of streams
streams easier to detect, by reporting one level fainter of the surface brightness limit, when the stream cannot be clearly distinguished by visual inspection at a certain level.

The morphology analysis shows 20-43\% shells (part of the cosmological class umbrella/shell) and 10-16\% circular shapes, some showing a clear loop around the host.

\begin{figure*}[h!]
\centering
\includegraphics[width=0.9\textwidth]{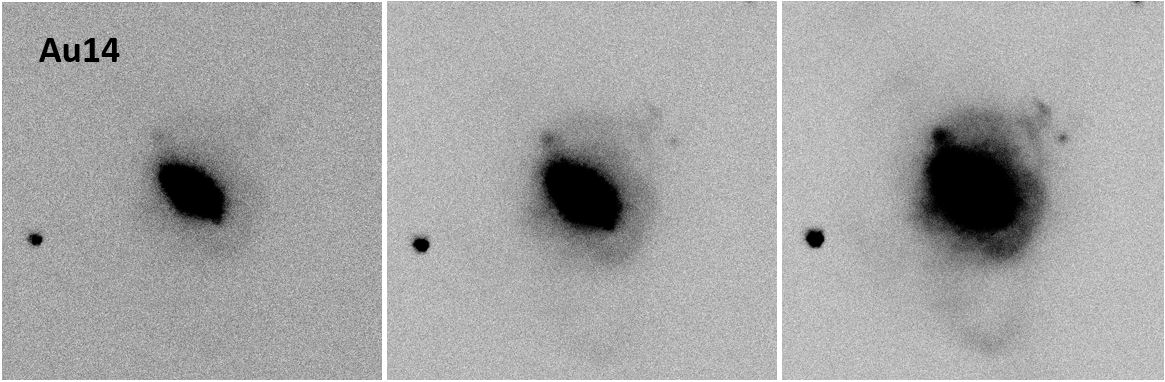}
\includegraphics[width=0.9\textwidth]{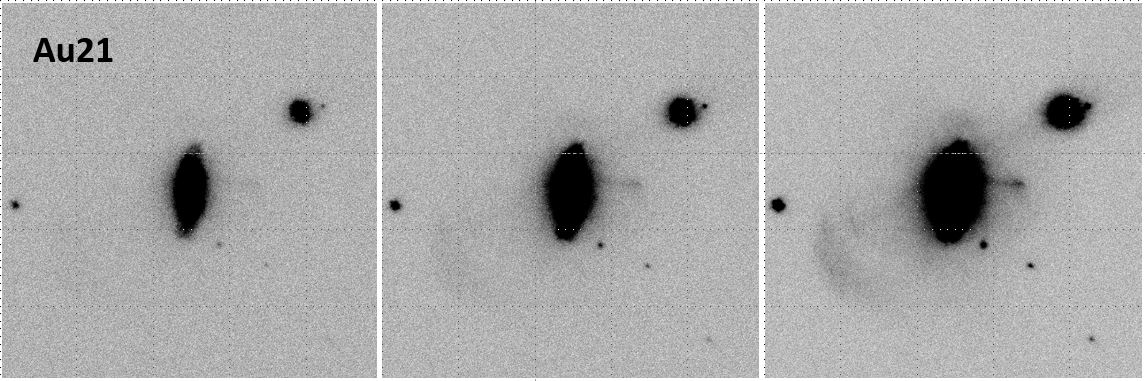}
\includegraphics[width=0.9\textwidth]{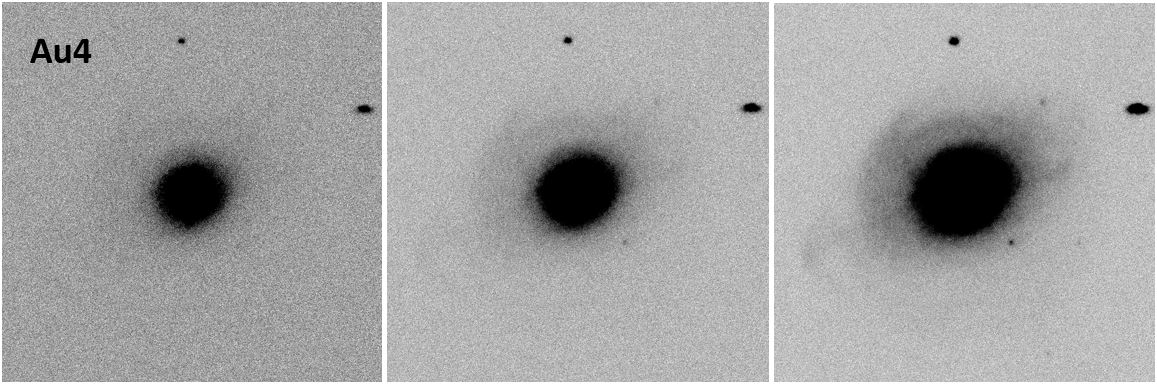}
\includegraphics[width=0.9\textwidth]{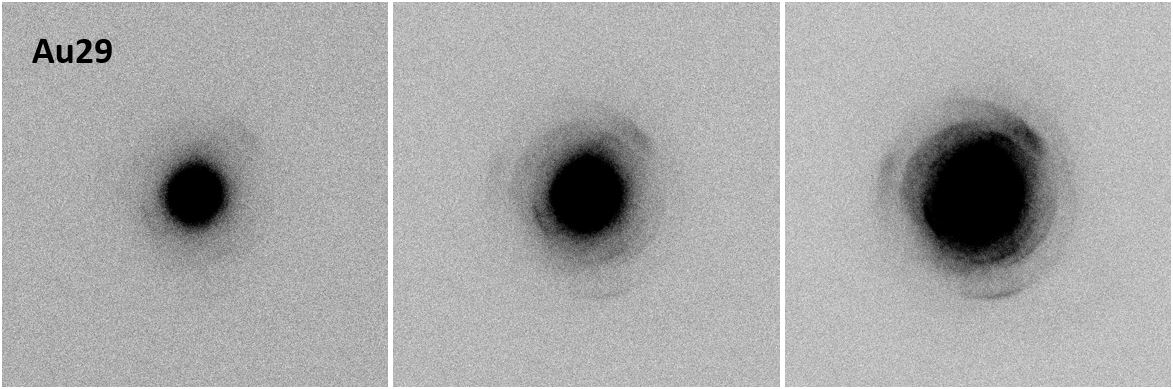}
  \caption{Examples of AURIGA mock images, Au4, Au14, Au21 and Au29 at different surface brightness limits. From left to right 29, 30 and 31 mag arcsec$^{-2}$. The streams appear clearer as the depth of the image increases from left to right. All images are $10\times10$ arcmin.} 
  \label{fig-aurigahaloexamples}
\end{figure*}

\subsection{Stellar Streams in the Auriga Simulation}
\label{sec:aurigasimulations}

We have available 30 Auriga zoom simulations of MW-mass halos
%halos (corresponding to 30 simulation snapshots 
at $z < 0.02$.
%)
%for visual inspection in search of stellar streams. 
Since 
%the results of analysing 30 halos 
this sample
may not be statistically representative, and in order to compare with the COCO and TNG50 simulations, three surface brightness maps have been generated from each halo, taking the three 
%x,y,z as lines-of-sight
orthogonal axes of the simulation coordinate system as lines of sight. None of the axes are aligned with face-on or edge-on directions so that overall the orientation of the galaxies is random. This process results in 
%having 
90 images, 
%and each 
each of which we combine with a range of sky backgrounds
between the SB limit of 25 and 34 mag arcsec$^{-2}$, at intervals of 1 mag arcsec$^{-2}$, thus yielding 900 mock images in total. Examples of halos with streams of different morphology can be seen in Figure \ref{fig-aurigahaloexamples}.

%\begin{figure}[h!]
%\centering
%\includegraphics[width=1.0\columnwidth]{Auriga_Stream-Detectability_240811.jpg}
% \caption{The solid line indicates detections of streams with a reasonable level of confidence. Detections of low surface brightness structures that cannot be confirmed to be streams are not included.}
%  \label{fig-AURIGAstreamdetection}
%\end{figure}

\begin{figure*}[h!]
\centering
\includegraphics[width=1.0\textwidth]{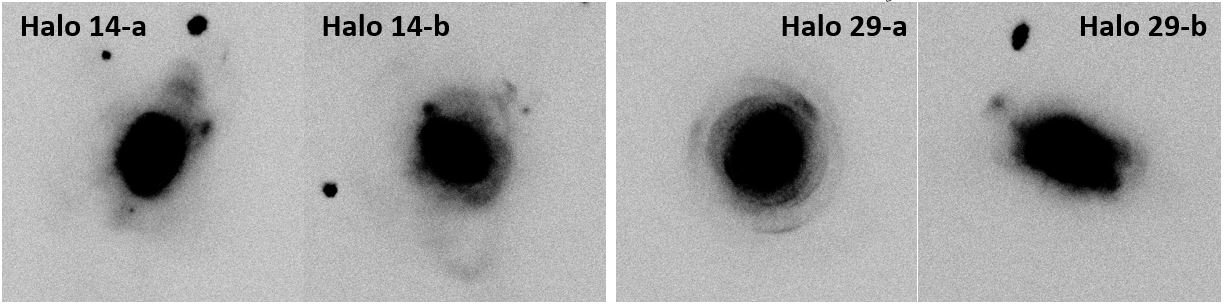}
 \caption{Example of the impact of the line-of-sight on morphology classification. \textit{left}: two views of Auriga halo 14 at a surface brightness limit of 31 mag / arcsec$^2$ from different line-of-sight. \textit{right}: the same views for Auriga halo 29  at a surface brightness limit of 31 mag / arcsec$^2$.}
  \label{fig-AURIGAloscomparison}
\end{figure*}

%and Figure \ref{fig-comparisonlsbdetectioncurve} for low surface b rightness structures that cannot be confirmed as streams.

\begin{figure}[h!]
\centering
\includegraphics[width=0.9\columnwidth]{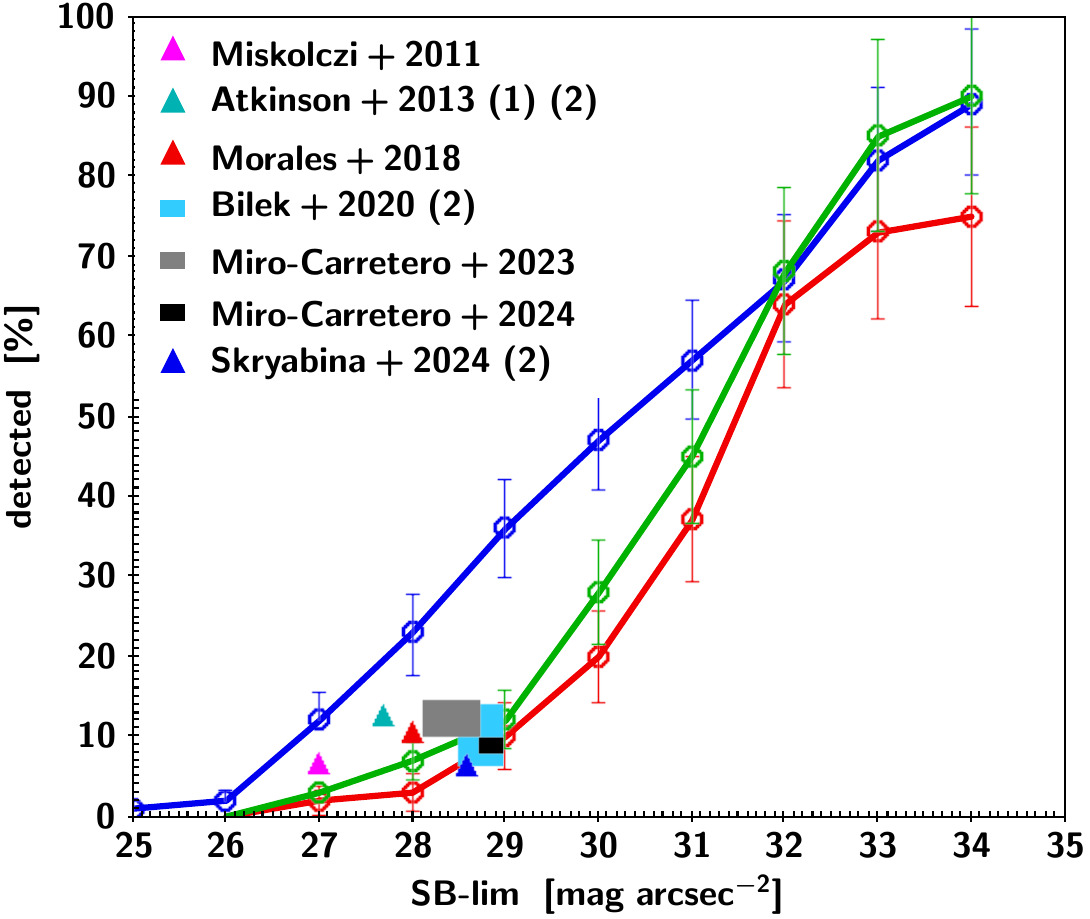}
  \caption{Comparison of the stellar stream detection rate as a function of the image surface brightness limit for TNG50 (\textit{red}), COCO (\textit{blue}) and AURIGA (\textit{green}) with the results of observations. (1) The calculation of the SB limit follows a different method (2) The detection rate has been derived for all LSB features, including streams.}
  \label{fig-comparisonstreamdetectioncurve}
\end{figure}

Figure \ref{fig-comparisondesonlyetectioncurve} shows the detection rate curve obtained for Auriga from visual inspection of the mock images, together with those obtained for COCO and TNG50, and the observational results of the DES sample. The figure shows the percentage of the sample for which at least one stream is detected for each of the 10 surface brightness limit levels analysed. The curve corresponds to detections for which there is a reasonable level of confidence that are streams. When we detect a LSB feature but have doubts about whether that LSB feature constitutes a stream, we take the conservative approach of not counting it towards the results included in the figure.
%to preserve purity of the findings and the detection is not classified as a stream, and not included in the graph.
Three of the halos (Au11 and Au30) seem to show an ongoing major merger between two galaxies of similar apparent size, clearly visible at a surface brightness limit of 26 and 27 mag arcsec$^{-2}$, respectively.
%, as the perceived size of the galaxies involved appears similar on the image. 
These three halos have been discarded, because we are looking for remnants of minor mergers only. LSB structures that could appear similar to streams at fainter surface brightness limits are not accounted for in the detectability curve.

The curve shows that there are no stream detections 
at a SB limit 26 mag arcsec$^{-2}$ or brighter. Between 26 and 29 mag arcsec$^{-2}$ there is an increase in detections with a gradient of around 4\% detection rate per 
mag arcsec$^{-2}$ the SB limit, reaching $\sim$ $12\%$ detection rate at 29 mag arcsec$^{-2}$. Between 29 and 31 mag arcsec$^{-2}$ the gradient is about 4 times steeper, reaching $\sim$ $34\%$ detection rate at 31 mag arcsec$^{-2}$. The gradient increases again to $\sim$ $25\%$ per mag arcsec$^{-2}$ increase in the SB limit between 31 and 32 mag arcsec$^{-2}$ reaching $\sim$ $65\%$ detection rate at this SB limit. Beyond this value, the gradient flattens.
We detect streams by visual inspection with a reasonable level of confidence in 95\% of the mock images up to a SB-limit of 34 mag arcsec$^{-2}$.
%however it is assessed that only 80\% of such faint structures can be assumed to be a stream with a reasonable level of confidence.
The percentage of streams detected in the AURIGA sample at a SB-limit corresponding to the average SB-limit for the DES sample in the $r$ band (28.65 mag arcsec$^{-2}$) is between 8\% and 13\% (corresponding to the SB-limit values for 28 and 29 mag arcsec$^{-2}$ in the curve, respectively).

The morphology analysis yields 41-62\% shells and 11-24\% circular shapes. However, the observed stream morphology is strongly dependent on the line-of-sight from which the streams is observed (Youdong et al. in prep.). As an example, Figure \ref{fig-AURIGAloscomparison} shows halos Au14 and Au29 seen from two different lines of sight. As can be appreciated in the images, looking at the same halo from these two different perspectives would suggest a different morphological classification.

\begin{table}
\centering
\small
{\caption{Percentages of the principal stream morphologies in observation and simulations.} 
\label{tab:morphology}

\begin{tabular}{lccc}

Sample & Shells & Circles & Total   \\
  &  $\%$ & $\%$   &  detections  \\

\hline\hline

DES	    &	$27-38$	&	$21-35$	&	$63$		\\
COCO	&	$70-90$	&	$2-3$	&	$108$		\\
TNG50	&	$20-43$	&	$10-16$	&	$79$		\\
AURIGA	&	$41-62$	&	$11-24$	&	$79$		\\

\hline

\end{tabular}
}
\begin{minipage}{9cm}
\vspace{0.3cm}
\small Note. Percentages the principal stream morphologies \textit{shell} and \textit{circle} identified in the the DES sample and in the mock images for the COCO, TNG50 and AURIGA simulations.The 4th  column indicates the total number of streams detected in the sample. Note that some images present more than one stream.
\end{minipage}
\end{table}

\section{Stream Photometry}
\label{sec:streamphotometry}

%This comparison between the observations and simulations includes also the results of the statistical analysis.
We have carried out a photometric analysis of the mock images generated from cosmological simulations though the process described in Section \ref{sec:mockimages}. We present here the results of the characterisation and photometric analysis of the mock-images generated from the COCO and TNG50 simulations and compare them with the ones obtained from the DES observations sample reported in \citep{miro-carretero2024}). 
In this paper, we restrict this photometric analysis to the two cosmological volume simulations, leaving a similar study of the more detailed higher-resolution Auriga simulations for future work (note also that due to the smaller number of Auriga simulations, such an analysis would have less statistical significance). To measure photometry parameters such as surface brightness and colours on the mock images the corresponding surface brightness maps have been generated for the SDSS $r$,$g$ and $z$ filters and transformed into counts images.

%The line of sight in the $z$ coordinate direction has been chosen, as the halos are not aligned with the volume coordinates, the chosen line of sight represents a random orientation of the galaxy and therefore it does not introduce an orientation bias of the system.  For the photometry analysis, the same smoothing method and length have been selected as for the stream detectability analysis in Section \ref{sec:streamdetection}. We have generated mock images from the surface brightness maps by transforming surface brightness readings to count readings, applying a zero point \apcedit{$ZP=22.5$~mag} as per DECam images. 

%Examples of mock-images analysed are shown in Figure \ref{fig-cocophotometryexamples}

%\begin{figure*}[h!]
%\centering
%\includegraphics[width=1.0\textwidth]{COCO-photometry-halo-287449-153.jpg}
%\includegraphics[width=1.0\textwidth]{COCO-photometry-halo-381188-153.jpg}
%\includegraphics[width=1.0\textwidth]{COCO-photometry-halo-465011-153.jpg}
%\includegraphics[width=1.0\textwidth]{COCO-photometry-halo-555817-153.jpg}
%\includegraphics[width=1.0\textwidth]{COCO-photometry-halo-410049-153.jpg}
%  \caption{Examples of COCO mock-images with streams processed with \textsc{Gnuastro/NoiseChisel}. \textit{left} $r$ band background subtracted image \textit{centre} zoom-in of the stream with a pixel size x 4 showing the apertures on which the phtotometry parameters are measured \textit{right} detection map} 
%  \label{fig-cocophotometryexamples}
%\end{figure*}

In order to compare the photometry of images with a similar surface brightness limit, the output of the simulations has been enhanced by superimposing a real DES sky background selected to have a surface brightness limit corresponding to the average surface brightness limit of the DES image sample with streams, namely 28.65 mag arcsec$^{-2}$ (see Section \ref{sec:mockimages}. As a result, only those mock images have been selected for the comparison, in which a stream was detected at a surface brightness limit of $\leq$ 28.65 This results in 21 mock-images from the COCO simulation and 8 mock images from the TNG50 simulation. Their photometry parameters are compared against those from 63 images of the DES sample with streams. This means that a statistically sound comparison is not possible, nevertheless the comparison is useful to see whether the cosmological simulations can reproduce real photometric observations, which requires realistic predictions of particle metallicity and age, which in turn is the result of the modelling of the physical processes involved.

The comparison between the mock-images photometry and the observations from the DES galaxy sample is carried out on the basis of the average stream surface brightness, average stream $(g-r)_{0}$ colour as well as the average distance of stream to the host centre. The resulting histograms are shown in Figures \ref{fig-cocodessb}, \ref{fig-cocodescolour} and \ref{fig-cocodesdistance}.

%Regarding stream morphology, there is a clear divergence between the DES observations and the COCO simulations. In the COCO mock images, 70-90\% of streams are Shells while in the DES sample images  only 27-38\% of streams are Shells. 

\begin{comment}

\begin{figure*}[h!]
\centering
\includegraphics[width=1.0\textwidth]{Fig-pixel-size_240903.jpg}
  \caption{Influence of pixel size on stream detection by visual inspection. \textit{left}: mock image of TNG50 halo 428177 with the DECam pixel size of 0.08 kpc and a surface brightness limit of 29 mag arcsec$^{-2}$ ; \textit{centre}: same image for a surface brightness limit of 32 mag arcsec$^{-2}$; \textit{right}: same image with a pixel size of 1 kpc and a surface brightness limit of 29 mag arcsec$^{-2}$.} 
  \label{fig-pixelsize}
\end{figure*}

\end{comment}

\section{Discussion}
\label{sec:discussion}

We now compare the predictions of the cosmological simulations with one another and with the observational data regarding frequency, detectability and morphology
of streams as well as their photometric properties.
Looking at Figure \ref{fig-comparisonstreamdetectioncurve}, the cosmological simulations all seem to predict that, for a surface brightness limit of 32 $\mathrm{mag \,arcsec^{-2}}$,
%a frequency of 
almost 70\% of galaxies in the mass range we study have one or more detectable streams.
%in the detection of streams around galaxies can be achieved.

%In \citet{johnston2008}, it is shown that the majority of tidal structures are expected to have a peak surface brightness fainter than $\sim$ 30 mag arcsec$^{-2}$.

%and that percentage does not seem to be dependent on the galaxy stellar mass for the range of stellar masses analysed in this work. 
However, the simulations show discrepancies with one another regarding 
%the 
detection rates 
%at 
for
surface brightness %levels below
limits brighter than
 $32\,\mathrm{mag \,arcsec^{-2}}$. 
In particular, the 
two hydrodynamical simulations,
%based on the SPH paradigm, 
TNG50 and Auriga, %while generally agreeing 
agree well with each other but predict a lower detectability rate than COCO.
%in that SB limit range. 
%Above  
For limits fainter than
$32\,\mathrm{mag \,arcsec^{-2}}$, however, TNG50
%falls short in detection rate with respect to COCO and Auriga, 
has a lower detection rate than COCO and Auriga (which agree well with each other, in this regime).
%that show similar values
The simulations also show discrepancies with one another regarding the morphology of the streams detected, as will be discussed further down in this section.

%Aside from 
The small differences in the stellar mass ranges
%in the simulations 
of the samples of galaxies drawn from the three simulations
(see Figure \ref{fig-histogramstelllarmass}) %, which
do not seem to play an important role in stream detectability for the range of stellar masses analysed in this work, as discussed in Section \ref{sec:cocosimulations}. We therefore speculate that these differences  %main reason for the differences 
%to be the physics modelling 
can be attributed mainly to the treatment of baryon physics in
%underlying 
the simulations (all the simulations use the same $N$-body treatment of gravitational dynamics and very similar cosmological parameters).  
Pinpointing specific explanations for these differences 
%the precise mechanisms in the physics treatment of both codes that account for the specific differences in the results 
%would 
will require an in-depth analysis, outside of the scope of this work.
However, since the most striking difference concerns the apparently greater number of relatively brighter streams detectable in the COCO particle tagging models, compared to the two hydrodynamical simulations, we speculate on why differences in the dynamical treatment of the baryons may give rise to this result. 

As discussed in \citet{cooper2017}, when comparing particle tagging simulations with hydrodynamical methods, even with identical initial conditions, it can be difficult to separate effects due to the dynamical approximation of particle tagging from the effects of different star formation models. In our case, the IllustrisTNG and COCO models have both been calibrated to observations of the galaxy mass and luminosity functions at $z=0$ \citep[in the case of COCO via the semi-analytic model of][]{lacey2016}, as well as other low-redshift data, notably the galaxy size--mass relation. The fundamental relationship between stellar mass and virial mass in both simulations agrees well with (for example) that inferred from galaxy abundance matching. Although the typical star formation histories of host galaxies in our sample, and their stream progenitors, may still differ in detail between the two simulations, we expect that they are broadly similar. This makes it more likely (although by no means certain) that the differences we observe are related to dynamical factors, rather than differences in how stars populate dark matter halos.

Two particularly relevant dynamical factors are neglected (by construction) in particle tagging models. First, the gravitational potentials of stream progenitors can be altered by the inflow and outflow of baryons associated with cooling and feedback; this may make satellites either more or less resilient to tidal stripping, depending on whether baryonic processes produce density cores or density cusps. Second, stars (and gas) could make a significant contribution to the central potential of the host halo, in particular through the formation of a massive baryonic disk.

\citet{cooper2017} explore the consequences of neglecting these factors when predicting satellite disruption (and hence stream and stellar halo formation) in particle tagging models. In the hydrodynamical model used in that work, massive satellites were found to disrupt somewhat earlier than realizations of the same systems in a particle tagging model\footnote{This of course depends on the detail of the hydrodynamical scheme and its subgrid recipes for star formation and feedback; the specific scheme used in IllustrisTNG was not examined by \citet{cooper2017}. A detailed case study of a single bright stream relevant to this discussion is given in Appendix A of that paper.}.
This could explain why more streams are visible at surface brightness limits of 
$\lesssim 28$ mag arcsec$^{-2}$
%$\lesssim 28\,\mathrm{mag arcsec^{-2}}$
%, streams are visible in DMO simulations that cannot be seen in SPH simulations, 
in COCO compared to IllustrisTNG; if satellites are disrupted earlier in IllustrisTNG, their streams may have more time to phase mix, lowering their surface brightness. 
%robably because they have been diffused.
%This phenomenon could 
A similar argument could also explain the greater abundance of shells detected in COCO. Shells originate from satellites on radial orbits \citep{newberg2016}.  
%that, as we have seen above, are subject to different interaction mechanisms with the host. 
The relative fraction of radial and circular orbits may differ between IllustrisTNG and COCO, because the evolution of the progenitor's orbit during pericentric passages (due to exchange of angular momentum with the host) may be significantly different with and without a massive baryonic disk.
%The fact that the COCO host galaxies do not have a disk also leads to differences in the interaction between the satellite and the host, resulting in divergent stream dispersion mechanisms that will affect the stream's surface brightness.
%We speculate that the differences in the interaction between the stream and the host during the infall are responsible for the difference in stream detectability results between COCO simulations and SPH based simulations, such as the TNG50 and Auriga simulations. 
A related effect is explored by \citet{valenzuela2024},
who study the relation between the rotation of the hosts and the presence and morphology of tidal features
%is analysed. 
One of the conclusions of that work is that shells appear more frequently in slow rotating hosts than in fast rotating hosts. This hypothesis could be tested in future work. More generally, further exploration of these differences particle tagging and hydrodynamical simulations could help to understand how observations of streams can constrain the nature of the gravitational potentials of stream progenitors and their host galaxies.

%Although the host rotation velocities in the simulations have not been compared, this could be a further reason for the overabundance of shells in the COCO simulations.  

Turning now to comparison between the simulations and the observational data, $8.7\pm1.1\%$ of galaxies in the DES sample have detectable streams at an average $r$-band surface brightness limit of 28.65 mag  arcsec$^{-2}$.
%Regarding the comparison of the predictions of the cosmological simulations and the observations on detectability, the percentage of streams found in the DES sample at an average $r$-band SB-limit
%in the $r$ pass band 
%of 28.65 mag  arcsec$^{-2}$ is 8.7 $\pm$ 1.1 \%. 
This seems to match very well with the predictions of TNG50 and Auriga, 
and is lower than the prediction from COCO.
%(as explained in the preceding paragraphs). 

%In contrast, the detection level inspecting the mock-images corresponding to that SB-limit value is above 20\%, as shown in the graph. Note that the  stream detection in the DES images, as reported in \citet{miro-carretero2024} covers a wider mass range.

% This good agreement is despite the fact that the stellar mass range for the TNG50 mock images is on the high side of the DES sample stellar mass distribution (from +1 to +2 $\sigma$). 
 
%Although the 
Stream morphology is not an observable that can be used reliably to constrain simulations, because it is strongly dependent on the line of sight along
which the stream is observed. Nevertheless, comparison of the 
%overall morphology figures 
fraction of streams with different morphologies is interesting because it could be related to the 
%specific modelling and simulation approaches. 
same dynamical differences between simulations methods that give rise to differences in stream abundance.
%The comparison of the TNG50 detection results with the DES sample regarding morphology is reasonable; TNG50 yields 
20-43\% of streams in the TNG50 sample are shells, which is within the range found in the DES sample, and 10-16\% have circular morphology, which is also not too far from the observations. 
As noted above, there is a significant discrepancy in the relative fractions of different stream morphologies %of the  streams detected
between observations and the COCO simulations. In the COCO mock images, 70-90\% of streams are identified as having Umbrella/Shell morphology, while in the DES sample,  
only 27-38\% of streams are identified as such. The COCO mock images show only 2-3 \% of streams with circular morphologies, compared to 21-35 \% in the DES images.
Auriga predicts 41-62\% of streams with Umbrella/Shell %stream
morphologies, significantly above the fraction in the DES sample, and 10-24\% of stream with circular morphology, somewhat below the DES figures.
Table \ref{tab:morphology} summarises the stream morphology findings for the different samples derived from the total number of streams detected. Note that more than one stream is detected in some of the halos. 
It is beyond the scope of this work to analyse the reasons for these discrepancies in 
stream morphology; as noted above, further work would be a valuable contribution to understanding the relationship between the observable properties of streams and the galaxy formation process as a whole.
%but we consider important to follow up on this so as to understand what are the physical or numerical parameters that influence the stream morphology. 

Regarding the stream photometry measurements, the stream average surface brightness range in the COCO simulation matches generally well the observations in the DES stream sample, though the distribution is skewed towards the brighter end of the range (Figure \ref{fig-cocodessb}). This is consistent with the abundance of shell-shaped streams, typically brighter than other stream morphologies, as confirmed by the observations. The range of the stream average surface brightness in the TNG50 simulation falls pretty much within the central region of the DES sample range.

The stream $(g-r)_{0}$ colour distribution of the COCO simulations also matches generally well the observations. The mean value of the COCO simulation distribution is 0.54$\pm$0.12 mag , versus 0.57$\pm$0.14 mag for the DES observations sample. However, the simulations are slightly skewed towards the blue end as can be seen in Figure \ref{fig-cocodescolour}. The TNG50 simulation, though not statistically significant, as we have only the value for 8 streams, shows a $(g-r)_{0}$ colour distribution spanning almost the full range covered by the DES observations.

Consistent with the overabundance of shell stream morphology in the COCO simulations, the average distance of the streams to the host galaxy is between 10 and 30 kpc, thus covering the lower end of the DES sample distance range, see Figure \ref{fig-cocodesdistance}. The streams from the TNG50 simulation are at an average distance between 20 and 50 kpc from the host centre, covering the central part of the distance range observed in the DES sample.

\subsection{Comparison with Previous Work}
\label{subsec:comparisonpreviouswork}

We present the comparison with other observations of streams as far as reported in the literature:

\begin{itemize}

\item \citet{miskolczi2011} reports a frequency of 6-19\% at a SB-limit of $\sim$ 27 mag arcsec$^{-2}$. We consider the lowest value of 6\% as the most reliable for maximising purity in the detection of streams (see Figure \ref{fig-comparisonstreamdetectioncurve}).

\item \citet{atkinson2013} reports a LSB feature (a superset of tidal streams) detection frequency of 12-18\% at an average SB-limit of $\sim$ 27.7 mag arcsec$^{-2}$ for the g-band. Note that the $g$-band in the DES sample is fainter in average than the $r$-band by $\sim$ 0.5 mag arcsec$^{-2}$. It is also to be noted that the calculation of the SB limit in that paper is based on 1$\sigma$ of the noise variation in apertures of 1.2 arcsec$^{2}$ placed on empty regions of the images and therefore differs from the one applied in this work [$3\sigma$, 100 arcsec$^{2}$] that follows the standard proposed in \citet{roman2020}. 

\item \citet{morales2018} reports a frequency of $\sim$ 10\% at a SB limit of 28 mag arcsec$^{-2}$.

\item \citet{bilek2020} reports 15\% of tidal features, including 5$\pm$2\% for streams and 5$\pm$2\% shells (which we consider together in our curve) at a SB limit of 28.5 - 29  mag arcsec$^{-2}$ for the $g$-band. Note that the $g$-band in the DES sample is fainter in average as the $r$-band by $\sim$ 0.5 mag arcsec$^{-2}$.

\item \citet{sola2022} have characterised the morphology of more than 350 low surface brightness structures up to a distance of 42 Mpc through annotation of images from the Canada--France Imaging Survey (CFIS2) and the Mass Assembly of early-Type gaLAxies with their fine Structures survey \citep[MATLAS,][]{duc2020,bilek2020}. They obtained 84 annotations for streams and 260 for shells, but out of these figures alone it is not possible to derive detectability figures for the brightest streams and shells, as there could be several of them in one image.

\item \citet{miro-carretero2023} report on the search of stellar tidal streams in DESI Legacy Survey images of MW-like galaxies, at distances between 25 and 40 Mpc, from the SAGA II galaxy sample \citep{geha2017,mao2021}. Applying the same detection and characterisation methods as in this work, their statistical analysis yields a stream detection frequency of 12.2 $\pm$ 2.4\% at a SB limit of 28.40 mag arcsec$^{-2}$. 

\item \citet{rutherford2024} use deep imaging from the Subaru-Hyper Suprime Cam Wide data to search for tidal features in massive [$\log_{10} M_{\star} / M_{\odot} > 10$] early-type galaxies (ETGs) in the SAMI Galaxy Survey. They report a tidal feature detection rate of 31 $\pm$2 \% at a surface brightness limit of 27 $\pm$0.5 mag arcsec$^{-2}$ for the $r$-band. They calculate the surface brightness limit following the same approach as \citet{atkinson2013}, thus different to our method [$3\sigma$, 100 $arcsec^2$]. For comparison, our method applied to a test Subaru HSC image yields a surface brightness limit of 29.79 mag arcsec$^{-2}$.

\item In \citet{skryabina2024} the results of visual inspection of a sample of 838 edge-on  galaxies using images from three surveys: SDSS Strip-82, Subaru HSC and DESI (DECals, MzLS,BASS) are presented. In total, 49 \textit{tidal features} out of 838 images are reported, equivalent to a frequency of $5.8\%$ at a SB limit of 28.60 mag arcsec$^{-2}$. In that study, tidal features include also disc deformations and tidal tails, typical of major mergers.

\end{itemize}

%\subsection{Comparison with Cosmological Simulations}
%\label{comparisonsimulations}

Also a number of papers in the literature report the frequency of tidal features (including streams and shells) in cosmological simulations:

\begin{itemize}

\item \citet{martin2022} reports on detection of tidal features inspecting mock-images produced using the NEWHORIZON cosmological simulations. Through production of surface brightness maps at different surface brightness limits, they predict the fraction of tidal features that can be expected to be detected at different limiting surface brightnesses.
%This is carried out on a sample of 37 simulated objects at redshifts  z = 0.2, 0.4, 0.6, and 0.8 spanning a stellar mass range of $10^{9.5} M_{\odot}$ < $M_{\star}$ < $10^{11.5} M_{\odot}$.
In this study, tidal features comprise: (i) Stellar streams, (ii) Tidal tails, (iii)
Asymmetric stellar halos, (iv) Shells  (v) Tidal bridges, (vi) Merger remnants (vii)
Double nuclei, of which only (i) and (iv) are clearly of accreted nature.
For a surface brightness limit of 35 mag arcsec$^{-2}$), expert classifiers were able to identify specific tidal features in close to 100 per cent of galaxies ($M_{\star}$ > $10^{9.5} M_{\odot}$), in agreement with our results.
For the range of stellar mass of the DES sample, $\log_{10} M_{\star}/M_{\odot} = 10$, they report $\sim$ 25\% detection rate for streams and shells together at a surface brightness limit of 30 mag arcsec$^{-2}$ in the $r$-band. The detection rate becomes 35\% for a surface brightness limit of 31 mag arcsec$^{-2}$. These figures match very well our predictions with Auriga and TNG50.

%At low masses ($M_{\star}$ < $10^{10} M_{\odot}$), almost no galaxies are expected to exhibit visible tidal features.
%At very faint limiting surface brightnesses ( μlim r (3 σ, 10 arcsec ×10 arcsec ) = 35 mag arcsec$^{-2}$), expert classifiers were able to identify specific tidal features in close to 100 per cent of galaxies ($M_{\star}$ > $10^{9.5} M_{\odot}$).
%In all, close to 100 per cent of galaxies exhibit some kind of distinct tidal feature (i.e. not just asymmetries) at μlim r (3 σ, 10 arcsec ×10 arcsec ) = 35 mag arcsec −2 regardless of mass. Ho we ver, this number falls fairly significantly as more realistic limiting surface brightnesses are considered. 

\item In \citet{khalid2024}, the results of identification and classification of \textit{tidal features}t in LSST-like mock-images from cosmological simulations are reported. Four sets of hydrodynamical cosmological simulations are used (\textsc{NewHorizons}, \textsc{EAGLE}, \textsc{IllustrisTNG} and \textsc{Magneticum}). The frequency of \textit{tidal features}, expressed in fractions of the total number of images, is between 0.32 and 0.40 showing consistency across the different simulations. \textit{Tidal Features} comprise streams/tails, shells, Plumes or Asymmetric Stellar Halos and Double Nuclei. Looking only at streams and shells, the percentage of detections varies between 5-15\% depending on the level of confidence, for a SB limit of 30.3 mag arcsec$^{-2}$ in the $r$-band.

\item In \citet{vera-casanova2022}, the authors report the results of inspecting surface brightness maps generated from 30 Auriga cosmological simulations \citep{grand2017} of MW-like galaxies looking for the brightest streams. These simulations are the same we have applied in this work. They report that no streams have been detected in images with a surface brightness limit brighter than 25 mag arcsec$^{-2}$ and that the stream detection frequency increases significantly between 28 and 29 mag arcsec$^{-2}$. For a surface brightness limit of 28.50 mag arcsec$^{-2}$ in the $r$-band they report a detection rate of $\sim$ 18 - 30\%. This is far higher than our results show with the same Auriga simulations and we believe the reason is the different pixel size of the mock-images. While we use the pixel size of the DECam instrument, 0.262 arcsec, equivalent to 0.08 kpc at 70 Mpc distance, in \cite{vera-casanova2022} they use a pixel size of 1 kpc. This is relevant for the visual inspection of counts images, as it has an impact on the S/N per pixel and could account for a significant difference in the surface brightness limit at which the streams are detected.
Rebinning the DECam images to a 1 kpc pixel size shows that this difference exceeds 2 mag arcsec$^{-2}$.
%2 mag  arcsec$^{-2}$  for the same detection rate. This is shown in Figure \ref{fig-pixelsize} where the DECam image has been warped to the 1 kpc pixel size (\textbf{figure to be added})

\item \citet{valenzuela2024} use the \textit{Magneticum} Box4 hydrodynamical cosmological simulations to detect tidal features (streams, shells and tidal tails) and connect their morphology to the internal kinematics of their host galaxies. In their Table 1 they present the fraction of galaxies with the different types of tidal features for host galaxies with $M_{\star} \geq 10^{11} M_{\odot}$. Looking at their Figure 3, the fraction of shells and streams together is $\sim$ 10$\%$ for galaxies with $10^{10} M_{\odot} < M_{\star} < 10^{11} M_{\odot}$, comparable with the stellar mass range of our simulations, at a surface brightness limit of 28.5 - 29 mag arcsec$^{-2}$. This results are similar to the ones obtained in this work.

\end{itemize}

\subsection{Caveats}
\label{subsec:caveats}

The comparison of our stream observability results with those of other surveys and cosmological simulations is not always straightforward, due to: i) the range of halo mass and stellar mass is not always the same, although our analysis of stream observability does not reveal a significant correlation of the stream observation rate with the host stellar mass within the range of stellar masses considered in this work; ii) the method applied for the  visual inspection of the images, some of the previous works seem to be based on the inspection of surface brightness maps with cut-offs of the surface brightness limit, while we inspect count images, where the pixel size bears an influence on the S/N per pixel and thereby on the detectability of streams;  iii) the different ways of calculating the surface brightness limits of the images by the different authors, as pointed out in the previous Section and iv) the different classification schemes used for the LSB features and their meaning, e.g. streams, shells or tidal tails used by the different authors.

The observability result obtained with mock-images can only be seen as an indicative reference for predicting the stream frequency to be met by present and future surveys for a number of reasons. First, this result has been obtained with an idealised flat image background, while in real-life observations the background is often not sufficiently flat, preventing the detection to reach the theoretical surface brightness limit, as explained in \citet{miro-carretero2024} regarding DES images. As mentioned earlier, for surface brightness limits much fainter than those of the DES sample, the confusion of sources and possibly cirri will become more significant and the synthetic background will be less representative of the real one, making detection more difficult. Second, this prediction is dependent on the modelling assumptions underlying the cosmological simulations. Nevertheless, the fact that hydrodynamic simulations provide very similar results with one another and match also the results from observations of the DES sample, as well as previous surveys and simulation analysis, provides confidence in the predictions.

Regarding the method of visual inspection of images, it is clear that the human factor plays a role in the results of the detection. However, the confidence in the method can be increased by having a team with experience in searching specifically for streams in real observations, and a systematic and rigorous method to proceed. While different scientists may come to different conclusions in specific cases, overall, in a large survey, these differences may not have a significant influence in the global results. Furthermore the inspection of mock-images with a flat background (as seen in Figures) does not leave a large margin of interpretation regarding the presence of tidal features. It is in the classification of these features where the divergences are more likely to appear. In any case, as discussed above, the morphology is a weak observable  because its perception is dependent 
%of
on
the line of sight. In the absence of a mature automatic detection method for extragalactic streams, visual inspection remains the state-of-the-art used by all the works in this domain reported in the literature.

\section{Conclusions and Outlook}
\label{sec:conclusions}

From the results obtained from comparing the stream frequency, characteristics and photometry of the DES galaxy sample observations with cosmological simulations, we can conclude that overall the methods applied here work well and provide a valid reference for the analysis of stellar streams.

Generally the predictions of the simulations are in agreement with the results of the   
analysis carried out on the DES sample, following the same approach to visual inspection as the present work, used as a reference \citep{miro-carretero2024}, and with previous work reported in the literature, as presented in Section \ref{subsec:comparisonpreviouswork}.
This provides a degree of confidence in the simulation predictions regarding detection of streams in future surveys at surface brightness limits for which we do not have observations today. The cosmological simulations we have have analysed here predict that, in the absence of a confusion limit due to background and foreground sources, and with a pixel size similar to the one of the DECam instrument, a frequency of almost 70\% in the detection of streams around galaxies can be achieved for a surface brightness limit of 32 mag $\mathrm{arcsec}^{-2}$. This prediction can be extrapolated to other observations taking into account the effect of the pixel size on the S/N.

Nevertheless, there are some noticeable differences in the stream morphologies between observations and simulations and between simulations themselves, that should be further analysed in order to understand their origin and 
%\apcedit{connection to assumptions made by the models.} 
be able to evolve the specific simulation resolution and physics modelling required for stream analysis where needed.

%in Overall the production of mock-images from cosmological simulations seems to provide a realistic reference for comparison with observations of streams, 

%albeit a certain lack of diversity in the morphology of the streams with respect to the observations. 
 
We present a method for comparison of stream observations with cosmological simulations based on novel tools for generation of mock images and measurement of photometry parameters. 
When inspecting the mock-images, we follow exactly the same approach and apply the same criteria, with the same team, as we did for inspecting the DES observational image sample, and reported in \citet{miro-carretero2024}.

In this work, we only compare with the output of the simulations at $z=0$.
%only a snapshot of the simulations output is analysed. Tapping into the simulation internals and following 
Exploring the evolution of streams from high redshift to the current time in the simulations could provide more insight into the history, mass ratio and kinematics of the preceding mergers,  allowing to compare the simulated reality with the conclusions of the visual inspection. 
This may also help to understand the differences in stream morphology that have been highlighted in this work.

Overall, the results of our work indicate that surveys with an instrument of similar characteristics and pixel-size as DECam and, reaching a surface brightness limit fainter than 31 mag arcsec$^{-2}$ would be required to attain a stellar tidal stream detection rate of at least 50\%, and thereby test the predictions of the $\Lambda$CDM model as implemented by state-of-the-art cosmological simulations.

%The results of the comparison between the findings of the statistical analysis versus the results of the cosmological simulations have been analysed and conclusi morphology ns are summarised here.

%Conclusions will be drawn on whether the potential differences identified between the two would allow to place constrains on the models on which the N-body simulations are based. If the conclusions so allow, new constraints on theoretical models of galaxy formation will be formulated on the basis of the analysis results

\begin{acknowledgements}

         JMC thanks the Leiden Observatory for hosting and providing computer infrastructure and facilities for carrying out part of this work, as well as the Universidad Complutense de Madrid for providing computer infrastructure used in this work.
         JMC thanks Yves Revaz for support in the use of the pNbody tool.

        JMC and MAGF acknowledge financial support from the Spanish Ministry of Science and Innovation through the project PID2022-138896NB-C55

         DMD acknowledges the grant CNS2022-136017 funding by MICIU/AEI /10.13039/501100011033 and the European Union NextGenerationEU/PRTR and finantial support from the Severo Ochoa Grant CEX2021-001131-S funded by MCIN/AEI/10.13039/501100011033 and project (PDI2020-114581GB-C21/ AEI / 10.13039/501100011033). 

        APC acknowledges support from a Taiwan Ministry of Education Yushan Fellowship and Taiwan National Science and Technology Council grants 112-2112-M-007-017 and 113-2112-M-007-009. 

        SRF acknowledge financial support from the Spanish Ministry of Economy and Competitiveness (MINECO) under grant numbers AYA2016-75808-R, AYA2017-90589-REDT, PID2021-123417OB-I00 and S2018/NMT-429. 
        
        For this work we have used GNU Astronomy Utilities (Gnuastro, ascl.net/1801.009) versions $0.17$, $0.18$ and $0.20$. Work on Gnuastro has been funded by the Japanese MEXT scholarship and its Grant-in-Aid for Scientific Research (21244012, 24253003), the European Research Council (ERC) advanced grant 339659-MUSICOS, and from the Spanish Ministry of Economy and Competitiveness (MINECO) under grant number AYA2016-76219-P.

        MA acknowledges the financial support from the Spanish Ministry of Science and Innovation and the European Union - NextGenerationEU through the Recovery and Resilience Facility project ICTS-MRR-2021-03-CEFCA and the grant PID2021-124918NA-C43.

        This work used high-performance computing facilities operated by the Center for Informatics and Computation in Astronomy (CICA) at National Tsing Hua University. This equipment was funded by the Ministry of Education of Taiwan, the Ministry of Science and Technology of Taiwan, and National Tsing Hua University.

        This work is supported by the National Science Center, Poland under Agreement No.  2020/39/B/ST9/03494.

        SB is supported by the UK Research and Innovation (UKRI) Future Leaders Fellowship (grant number MR/V023381/1).

        CSF acknowledge STFC Consolidated Grant ST/X001075/1 and support from the European Research Council through ERC Advanced Investigator grant, DMIDAS [GA 786910] to CSF. This work used the DiRAC@Durham facility managed by the Institute for Computational Cosmology on behalf of the STFC DiRAC HPC Facility (www.dirac.ac.uk). The equip- ment was funded by BEIS capital funding via STFC capi- tal grants ST/K00042X/1, ST/P002293/1 and ST/R002371/1, Durham University and STFC operations grant ST/R000832/1.DiRAC is part of the National e-Infrastructure.

\end{acknowledgements}

\begin{appendix}

\section{Mock-Images Photometry}
\label{appendix:mockimagesphotometry}

\begin{figure}[h!]
\centering
\includegraphics[width=0.8\columnwidth]{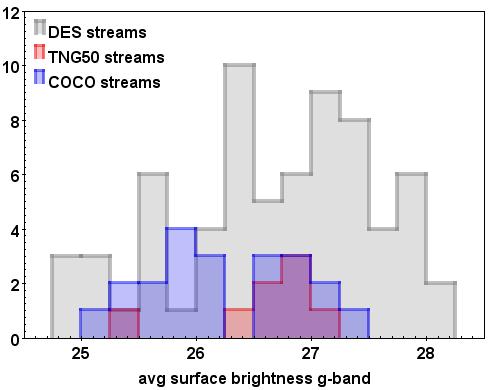}
\includegraphics[width=0.8\columnwidth]{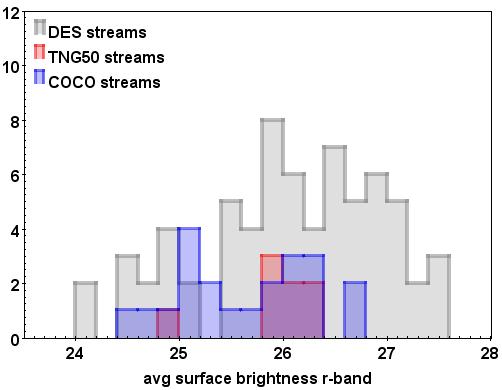}
\includegraphics[width=0.8\columnwidth]{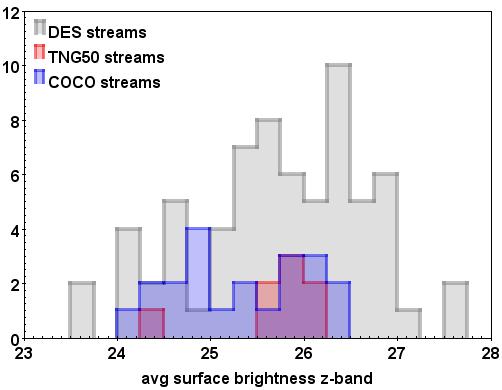}
  \caption{Histogram of the distribution of average surface brightness measured on the stream for the $g$ (top) $r$ (centre) and $z$ (bottom) bands.} 
  \label{fig-cocodessb}
\end{figure}

\begin{figure}[h!]
\centering
\includegraphics[width=0.8\columnwidth]{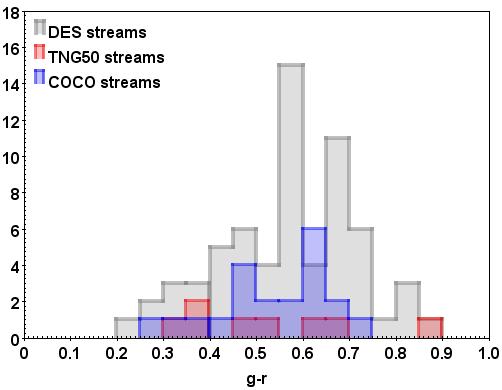}
  \caption{Histogram of the distribution of average $(g-r)_{0}$ colour  measured on the stream. } 
  \label{fig-cocodescolour}
\end{figure}

\begin{figure}[h!]
\centering
\includegraphics[width=0.8\columnwidth]{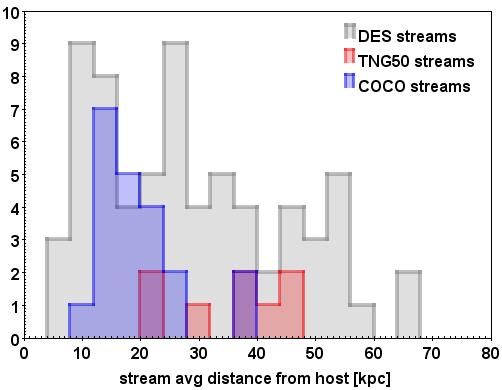}
  \caption{Histogram of the distribution of average distance of the streams to the centre of the host galaxy. } 
  \label{fig-cocodesdistance}
\end{figure}

\end{appendix}

\end{document}